\documentclass[nofootinbib,superscriptaddress,showpacs,showkeys]{revtex4}

\usepackage{graphicx,epsfig}
\usepackage{amsfonts,amsmath,amssymb,amsthm,amscd}
\usepackage[T2A]{fontenc}
\usepackage{color}

\begin{document}

\title{Enhancing proton acceleration by using composite targets}

\author{S. S. Bulanov}
\affiliation{Lawrence Berkeley National Laboratory, Berkeley, California 94720, USA}

\author{E. Esarey}
\affiliation{Lawrence Berkeley National Laboratory, Berkeley, California 94720, USA}

\author{C. B. Schroeder}
\affiliation{Lawrence Berkeley National Laboratory, Berkeley, California 94720, USA}

\author{S. V. Bulanov}
\affiliation{QuBS, Japan Atomic Energy Agency, Kizugawa, Kyoto, 619-0215, Japan}
\affiliation{A. M. Prokhorov Institute of General Physics RAS, Moscow, 119991, Russia}

\author{T. Zh. Esirkepov}
\affiliation{QuBS, Japan Atomic Energy Agency, Kizugawa, Kyoto, 619-0215, Japan}

\author{M. Kando}
\affiliation{QuBS, Japan Atomic Energy Agency, Kizugawa, Kyoto, 619-0215, Japan}

\author{F. Pegoraro}
\affiliation{Physics Department, University of Pisa and Istituto Nazionale di Ottica, CNR, Pisa 56127, Italy}

\author{W. P. Leemans}
\affiliation{Lawrence Berkeley National Laboratory, Berkeley, California 94720, USA}
\affiliation{Physics Department, University of California, Berkeley, California 94720, USA}

\begin{abstract}
Efficient laser ion acceleration requires high laser intensities, which can only be obtained by tightly focusing laser radiation. In the radiation pressure acceleration regime, where the tightly focused laser driver leads to the appearance of the fundamental limit for the maximum attainable ion energy, this limit corresponds to the laser pulse group velocity as well as to another limit connected with the transverse expansion of the accelerated foil and consequent onset of the foil transparency. These limits can be relaxed by using composite targets, consisting of a thin foil followed by a near critical density slab. Such targets provide guiding of a laser pulse inside a self-generated channel and background electrons, being snowplowed by the pulse, compensate for the transverse expansion. The use of composite targets results in a significant increase in maximum ion energy, compared to a single foil target case.                 
\end{abstract}

\maketitle

\section{Introduction}

{\noindent}The laser-driven acceleration of charged particles is the main focus of many existing, being constructed, and projected high-intensity laser facilities \cite{Review,BELLA,ELI}. In particular, the acceleration of ions has attracted a significant  attention both theoretically and experimentally \cite{review}, due to the many potential applications of laser-accelerated ion beams, including but not limited to, fast ignition \cite{FI}, hadron therapy \cite{hadron therapy}, radiography of dense targets \cite{radiography}, injection into conventional accelerators \cite{injection}, and nuclear physics studies \cite{nuclear}.

Several years ago Particle-in-Cell (PIC) based computer simulations predicted that several hundred MeV or even GeV ion beams can be produced in laser-matter interactions using modern multi-TW or PW laser systems \cite{MeV protons}. Recently, several new acceleration schemes were proposed that offer the possibility of generation multi-GeV or TeV proton beams \cite{TeV protons, slow wave_PRL}. However such schemes would require next generation lasers with even higher peak intensity. There are several basic laser ion acceleration mechanisms that has been discussed in the literature: (i) Target Normal Sheath Acceleration (TNSA) \cite{TNSA}, (ii) Coulomb Explosion (CE) \cite{CE}, (iii) Radiation Pressure Acceleration (RPA) \cite{RPA}, (iv) Magnetic Vortex Acceleration (MVA) \cite{MVA_first,MVA,MVA_new}, and (v) the Shock Wave Acceleration (SWA) \cite{SWA}. 

TNSA is usually realized in the case of relatively thick targets and poor laser contrast, producing ion beams with Maxwellian-type energy spectra. In TNSA the laser launches electrons from the front of the target all the way through the target. As they leave the target from the back, those electrons establish a quasistatic longitudinal electric field that accelerates ions from the rear side of the target. Most of the experimental results on proton acceleration can be attributed to the TNSA scheme, with the peak proton energy around 70 MeV \cite{TNSA_exp}.

The CE acceleration works for high intensity pulses that are able to remove all the electrons from the irradiated spot, ideally without disturbing the ion core. The remaining ion core experiences CE due to the noncompensated positive charge. Usually a two layer (high Z/low Z) target is used to enhance the properties of CE accelerated ions. The low-Z ions are accelerated in the field of expanding high-Z layer producing a quasimonoenergetic spectrum, which is required for almost all laser driven ion beam applications. However extremely high intensity and high laser contrast, not only at the nanosecond scale but also at the picosecond scale make this scheme very challenging for experimental realization. 

The RPA mechanism comes into play when the laser is able to push the foil as a whole by its radiation pressure. It is the realization of the relativistic receding mirror concept \cite{RM review}. The role of a mirror is played by an ultra-thin solid density foil or by plasma density modulations emerging when the laser interacts with an extended under-critical density target, the so-called hole-boring RPA \cite{hole-boring RPA}. The laser pulse is reflected by a co-moving mirror. The problem of a plane EM wave reflection by a mirror moving with a relativistic velocity was considered by A. Einstein as an illustration of the Theory of Special Realtivity \cite{Einstein}. The frequency of the reflected radiation is shifted down by a factor of $4\gamma^2$, where $\gamma$ is the Lorentz factor of the mirror. Thus the energy transferred to the mirror is $(1-1/4\gamma^2)\mathcal{E}_L$, where $\mathcal{E}_L$ the energy of the laser pulse. For $\gamma\gg 1$ almost all laser energy is transferred to the foil, which makes this scheme very attractive in the ultrarelativistic limit \cite{RPA,COMREN,RPA_recent}. Recently, there were several papers published that claim the experimental observation of the onset of this regime of laser ion acceleration \cite{RPA exp}. 

The MVA regime is different from the other three regimes, and is realized when the laser interacts with a Near Critical Density (NCD) target (TNSA, CE, and RPA rely on the interaction with solid density ultra-thin foils). In MVA the laser pulse propagates in NCD plasma, it makes a channel in both electron and ion density. When the laser pulse exits the plasma it establishes strong longitudinal electric fields that can accelerate the ions which are pinched by the electrons in a thin filament along the laser propagation axis. In the case of long pulses, the acceleration of helium atoms up to 40 MeV from underdense plasma was observed at the VULCAN laser \cite{Willingale2006} and the acceleration of protons up to 50 MeV at Omega EP \cite{Willingale2011}. Also the experiments with short pulses and cluster jets show that 10-20 MeV per nucleon ions can be generated via such interaction \cite{Fukuda}. Like MVA the SWA regime employes an underdense plasma, where the laser launches a shock wave inside the target. The ions in the bulk of the target are reflected by this shock wave at twice the velocity of the shock. Ideally such mechanism should produce monoenergetic ions, since the reflection happens after the shock is detached from the laser pulse. However, the accelerated ion beam will interact with the background plasma and will be subject to two-stream and filamentation instabilities \cite{slow wave}, the latter determining the width of the accelerated ion spectra.

There are several mechanisms that, through either modification or combination of some of the basic regimes, enhance the maximum ion energy, number of accelerated ions, or improve their spectrum. For example, the Burn-out-Afterburner (BOA) \cite{BOA}, which employes an enhanced TNSA, and Directed Coulomb Explosion (DCE) \cite{DCE}, which is the combination of RPA and CE, are such composite mechanisms. There was also an extensive study connected with ion acceleration during the induced relativistic transparency of the target \cite{transparency}. Also the use of composite targets (low density/high density) was proposed in a number of papers to either inject the ions into accelerating fields or to enhance the interaction of the laser pulse with the high density part of the target \cite{TeV protons, slow wave_PRL, Macchi, Zepf}.  

In this paper we propose to enhance the maximum attainable energy of ions accelerated via RPA by utilizing composite targets, consisting of a solid density ultra-thin foil placed in front of the NCD slab. The RPA accelerated ions are subject to  a number of factors limiting their maximum energy. Typically RPA requires high laser intensities, which can only be obtained by tightly focusing laser radiation. In this case the finite spot size effects are important \cite{Dollar}. Tightly focused laser pulses have a group velocity less than the speed of light in vacuum, which would limit the maximum attainable ion energy to the energy corresponding to the group velocity \cite{slow wave,slow wave_PRL}. In the framework of the receding relativistic mirror concept, the energy transferred from the pulse propagating with the group velocity, $\beta_g$, smaller than the speed of light in vacuum to the mirror is $\Delta\mathcal{E}\simeq 2\gamma^2\beta(\beta_g-\beta)\mathcal{E}_L$. Thus, if $\beta=\beta_g$, then there is no interaction of the laser light with the target, and consequently no energy transfer. Moreover tightly focused laser pulses force the foil to expand in the transverse direction during the interaction, making it transparent for radiation and effectively terminating the acceleration. In the case of a composite target, the laser, after beginning to accelerate the foil, enters the NCD slab and generates a channel in electron and ion density. Inside this self-generated channel the laser pulse propagates at subluminal group velocity, $\beta_g<1$, which is a function of the laser pulse power, over the depletion length, which is also the acceleration length for the foil. For a tightly focused pulse, the transverse expansion of the foil is the main limiting factor for the acceleration. The acceleration distance for the expanding foil is substantially smaller than the depletion length in the NCD plasma. Thus the interaction of a laser pulse with a composite target leads to the enhancement of the maximum ion energy. 

The paper is organized as follows. In section 2 we review the equations of the foil motion under the action of the laser pulse radiation pressure. In section 3 we consider the effect of the transverse expansion of the target. We present the results of PIC simulations in section 4 and conclude in section 5. We discuss the properties of the laser pulse propagating in the NCD plasma and the energy of ions accelerated via the MVA mechanism in Appendices A and B respectively.          

\section{Radiation Pressure Acceleration}

{\noindent}In this section we review the basic properties of the RPA mechanism. The motion of the foil is  modeled by an interaction of an EM wave with a mirror described by the equations \cite{RPA, slow wave, optimized RPA}. The laser pulse radiation pressure is taken into account as a force proportional to the EM wave momentum, which in  turn is proportional to the EM wave Pointing vector, $\mathbf{S}=\mathbf{E}_L\times\mathbf{B}_L/4\pi$, where $E_L$ and $B_L$ are the electric and magnetic fields of the laser pulse. We assume $c=1$ throughout the paper. If we consider the RPA by a plane wave of a non-expanding opaque foil in the 1D approximation \cite{RPA}, then the momentum of the foil is 
\begin{equation}
p=\frac12\left[\left(1+W\right)-\frac{1}{\left(1+W\right)}\right].
\end{equation} 
Here $W=2\mathcal{F}_L/n_e l$,  
\begin{equation}
\mathcal{F}_L=\int\limits_0^\infty\frac{|E_L(\eta)|^2}{4\pi} d\eta
\end{equation}
is the laser pulse fluence (incident laser energy per unit area) and $n_e l$ is the areal density of the foil, with $n_e$ the electron foil density and $l$ the foil thickness. In the ultrarelativistic case the energy of the foil asymptotes to  \cite{RPA}
\begin{equation}
\gamma_{\beta_g=1}^{max}=\frac{\mathcal{F}_L}{n_e l}.
\end{equation}
The subscript $\beta_g=1$ indicates that the laser group velocity is equal to the speed of light in vacuum in this case.

In a more general case of an expanding target and a laser pulse with group velocity smaller than the speed of light in vacuum the equation of motion of the surface element of a mirror in the laboratory frame of reference can be written in the following form \cite{unlimited,slow wave}:
\begin{equation}
\frac{d\mathbf{p}}{dt}=\frac{\mathcal{P}}{n_e l}\mathbf{\nu},
\end{equation}  
where the force is proportional to the laser pressure and directed along the normal to the surface element, denoted by unit vector $\mathbf{\nu}$. The light pressure is determined from balancing the fluxes of the incident, reflected, and transmitted EM waves in the reference frame, where the particular surface element is at rest. In what follows we will refer to this reference frame as the moving frame or M-frame, and to the laboratory frame as the L-frame. In the M-frame the force acting on the foil is    
\begin{equation}
\mathbf{F}=(1+|\rho|^2-|\tau|^2) (\mathbf{S}\cdot\mathbf{\nu})\mathbf{\nu}.
\end{equation} 
The reflection coefficient, $\rho$, the transmission coefficient, $\tau$, and the absorption coefficient, $\alpha$, are related to each other through energy conservation: $|\rho|^2+|\tau|^2+|\alpha|^2=1$, and are determined from the solution of the wave equation for the EM wave interacting with a thin foil in the  M-frame \cite{Vshivkov}. 

The maximum attainable ion energy for RPA by laser pulses having their peak intensity and consequently peak fluence on-axis is also achieved by an on-axis surface element of the foil. For this element the laser propagtion direction coincides with the vector normal to the surface. For a circularly polarized EM wave with amplitude of the vector-potential, $A_0$, frequency $\omega$, and wave vector $\mathbf{k}$, propagating along the $x$ axis we have
\begin{equation}
\mathbf{S}=\omega k A_0^2\mathbf{e}_x,
\end{equation} 
where $A_0$ is the amplitude of the EM wave vector potential. Following the results of Ref. \cite{slow wave}, we assume that the EM wave travels with a group velocity $\beta_g=k/\omega$, which is the case for focused pulses and waves traveling inside some external guiding structure. In the boosted frame of reference, moving with velocity $\beta$, we obtain:
\begin{equation}
\overline{\omega}\overline{k}=\omega^2\frac{(\beta_g-\beta)(1-\beta\beta_g)}{1-\beta^2}.
\end{equation}  
Thus the equation of the on-axis surface element motion is 
\begin{equation}\label{eq of motion}
\frac{d\beta}{dt}=\kappa\beta_g(1-\beta^2)^{1/2}(\beta_g-\beta)(1-\beta\beta_g),
\end{equation}    
where time is measured in units of $\omega^{-1}$, and
\begin{equation}
\kappa=(2|\rho|^2+|\alpha|^2)\frac{\omega A_0^2}{4\pi n_e l m_i}=\frac{1}{2}(2|\rho|^2+|\alpha|^2)\frac{m_e}{m_i}\frac{a^2(\psi)}{\varepsilon_0}. 
\end{equation}
Here $a=eA/m_e$ is the dimensionless vector-potential, $\varepsilon_0=\pi (n_e l/n_{cr}\lambda)$ is the parameter governing the transparency of the thin solid density target \cite{Vshivkov}, $n_{cr}=m_e\omega^2/4\pi e^2$ is the critical plasma density, $e$ and $m_e$ are the electron charge and mass respectively, $\psi=(t-x/\beta_g)$ is the phase, and $m_i$ is the ion mass. The reflection coefficient is determined from the solution of the wave equation in the rest frame of the foil \cite{optimized RPA,Vshivkov}: 
\begin{equation}
\rho=\frac{\tilde{\epsilon}_0}{a_0}\left[\frac{b}{(1+b^2)^{1/2}}\right],
\end{equation}
\begin{equation}
b=\frac{1}{2^{1/2}}\left\{\left[(a_0^2-\tilde{\epsilon}_0^2-1)^2+4a_0^2\right]^{1/2}+(a_0^2-\tilde{\epsilon}_0^2-1)\right\}^{1/2},
\end{equation} 
where $\gamma=(1+p^2)^{1/2}$, and $\tilde{\epsilon}_0=\gamma\epsilon_0$, which accounts for the fact that the parameter $\epsilon_0$ is not a relativistic invariant \cite{optimized RPA}. We notice that the condition $a_0=\epsilon_0$ is considered to mark the optimal regime of ion acceleration \cite{optimal}. However this is true only in the case of non relativistic ion energies. In the relativistic case, the condition transforms into $a_0=\gamma\epsilon_0$ \cite{optimized RPA}. In what follows we assume total reflection and set $|\rho|=1$. For $\beta=\beta_g$, the right hand side of Eq. (8), i.e., the radiation pressure is equal to zero. This will result in the appearance of the maximum attainable velocity equal to $\beta_g$ for the ions accelerated by the RPA of an EM wave with $\beta_g<1$. The equation (\ref{eq of motion}) can be solved in quadratures. For the sake of brevity we assume that initially the foil was at rest, $\beta(0)=0$. This yields the following expression:
\begin{eqnarray}
\left\{\ln\frac{(1-\beta\beta_g+(1-\beta_g^2)^{1/2}(1-\beta^2)^{1/2}\beta_g}{(\beta_g-\beta)(1+(1-\beta_g^2)^{1/2})}-\beta_g\left[\arctan\frac{(1-\beta_g^2)^{1/2}(1-\beta^2)^{1/2}}{\beta_g-\beta} 
-\arccos \beta_g\right]\right\}  \nonumber\\ \nonumber\\
=\beta_g(1-\beta_g^2)^{3/2}K_\beta(t). \label{eq of motion_time},
\end{eqnarray} 
where $K_\beta(t)=\int\limits_0^t \kappa dt^\prime$. In order to get an idea of the foil's behavior as its velocity approaches $\beta_g$, we assume that the EM field is constant, then $K_\beta(t)=\kappa t$ and for $t\rightarrow\infty$ only the term with $\ln(\beta_g-\beta)$ survives. Thus we obtain the following asymptotic expression for the ion velocity:
\begin{equation}\label{group velocity limit}
\beta=\beta_g-\exp\left(-\beta_g(1-\beta_g^2)^{3/2}\kappa t\right). 
\end{equation}
From Eq. (\ref{group velocity limit}) one can see that the foil velocity approaches the laser group velocity exponentially, but never exceeds it. Thus the ion energy is limited by $\gamma_g=(1-\beta_g^2)^{-1/2}$ and the time needed to accelerate the foil to the energy approaching $\gamma_g$ scales logarithmically with $\gamma$:
\begin{equation}
t=\frac{\gamma_g^3}{\kappa\beta_g}\ln\left[\frac{2\gamma_g\gamma}{\gamma_g-\gamma}\right].
\end{equation}
We should mention here that, for finite duration laser pulses, the group velocity limit will manifest itself only for $\gamma_{\beta_g=1}^{max} > \gamma_g$, i.e., in the case where the maximum accelerated ion energy by the same pulse (but with $\beta_g=1$), would exceed the energy determined by the group velocity.

In Fig. 1 we present the numerical solution of Eq. (\ref{eq of motion_time}) for a Gaussian pulse with a duration of 27 fs (10 cycles), $\lambda=800$ nm wavelength, focal spot of $w_0=1.35\lambda$, which corresponds to an f-number of $f/D=1.5$, interacting with  a $0.25\lambda$ thick hydrogen foil with an electron density of $400n_{cr}$. The evolution of the maximum ion velocity (Fig. 1a) and maximum ion energy (Fig. 1b) are shown for three different values of the averaged laser power (1 PW, 1.8 PW, and 3.6 PW) and two values of group velocity $\beta_g=1$ and $\beta_g=0.986$ ($\gamma_g=6$, $w_0=1.35\lambda$, f/D=1.5 \cite{group velocity}). For P=1 PW, the evolution of ion velocity (and energy) is very similar for $\beta_g=1$ and $\beta_g<1$ cases, since $\gamma_{\beta_g=1}^{max}<\gamma_g$ and the effects group velocity are small. For P=1.8 PW, one can see a small difference between these two cases, which is mainly due to different dephasing between the pulse and the foil lengths. We will address this effect below. For P=3.6 PW, $\gamma_{\beta_g=1}^{max}>\gamma_g$, which leads to significant differences between $\beta_g=1$ and $\beta_g<1$ cases, as expected from the analytical analysis. The $\beta_g<1$ cases are limited by $\beta_g$ for velocity and $\gamma_g-1$ for energy, while the $\beta_g=1$ cases are only limited by the total energy stored in the laser pulse.

We notice that for P=1.8 PW, the $\beta_g<1$ case results in higher maximum ion velocity (energy). This is due to the fact that, when the foil reaches relativistic velocities, the dephasing between the foil and the laser plays an important role. The pulse with $\beta_g<1$ will interact with the foil longer, thus accelerating it to higher energies. If we rewrite Eq. (\ref{eq of motion}) in terms of the phase $\psi=(t-x/\beta_g)$, then
\begin{equation}\label{eq of motion_phase}
\frac{d\beta}{d\psi}=\kappa \beta_g^2(1-\beta^2)^{1/2}(1-\beta\beta_g).
\end{equation} 
This equation can also be solved in quadratures:
\begin{equation} \label{K(beta)}
\frac{1}{\beta_g(1-\beta_g^2)^{1/2}}\left\{\arctan\left[\frac{(1-\beta_g^2)^{1/2}(1-\beta^2)^{1/2}}{\beta_g-\beta}\right]-\arccos \beta_g\right\}=K_\beta(\psi),
\end{equation}
where $K_\beta(\psi)=\int\limits_0^\psi \kappa d\psi^\prime$. From Eq. (\ref{K(beta)}) we obtain for the surface element velocity:
\begin{equation}\label{beta(phi)}
\beta=\frac{\beta_g\sin^2R-(1-\beta_g^2)\cos R}{1-\beta_g^2 \cos^2 R},
\end{equation}
where 
\begin{equation}
R=K_\beta(\psi)\beta_g^2(1-\beta_g^2)^{1/2}+\arccos \beta_g. 
\end{equation}
Since $0\leq\beta\leq\beta_g$, the parameter $R$ is defined on the interval $\arccos\beta_g\leq R\leq \pi/2$. In order to compare the cases $\beta_g=1$ and $\beta_g<1$ we assume that $K_\beta(\psi)=\kappa\psi$ and write down the expression for the final velocity, in the case $\beta_g=1$, in the following form
\begin{equation}
\beta_0=\frac{\kappa\psi\left(2+\kappa\psi\right)}{2+2\kappa\psi+\left(\kappa\psi\right)^2}.
\end{equation}       
In the case of $\beta_g<1$ we use Eq. (\ref{K(beta)}), assume that $\beta_g\rightarrow 1$, expand the lefthand side around $\beta_g=1$, and solve with respect to $\beta$,
yielding
\begin{equation}
\beta=\beta_0+\frac{16}{3}\frac{\kappa\psi\left(1+\kappa\psi\right)}{\left[2+2\kappa\psi+\left(\kappa\psi\right)^2\right]^2}\left(1-\beta_g\right).
\end{equation}  
Thus one can see that, due to slower dephasing, the laser pulse with $\beta_g<1$ accelerates the foil to higher energies than the laser pulse with $\beta_g=1$. However this consideration is valid only for $\beta<\beta_g$. This dephasing is illustrated in Fig. 1c, where the evolution of the value of the field amplitude, $a_0(x)$, at the foil is shown. For the particular case of P=1.8 PW wee see that $\beta_g<1$ curve has higher values than the $\beta_g=1$ curve for an extended period of time. In the case of a 3.6 PW laser, the pulse becomes phase locked with the foil, \textit{i.e.,} it co-propagates with it, but can not transfer any amount of energy to the foil.

\begin{figure}[tbp]
\epsfxsize6cm\epsffile{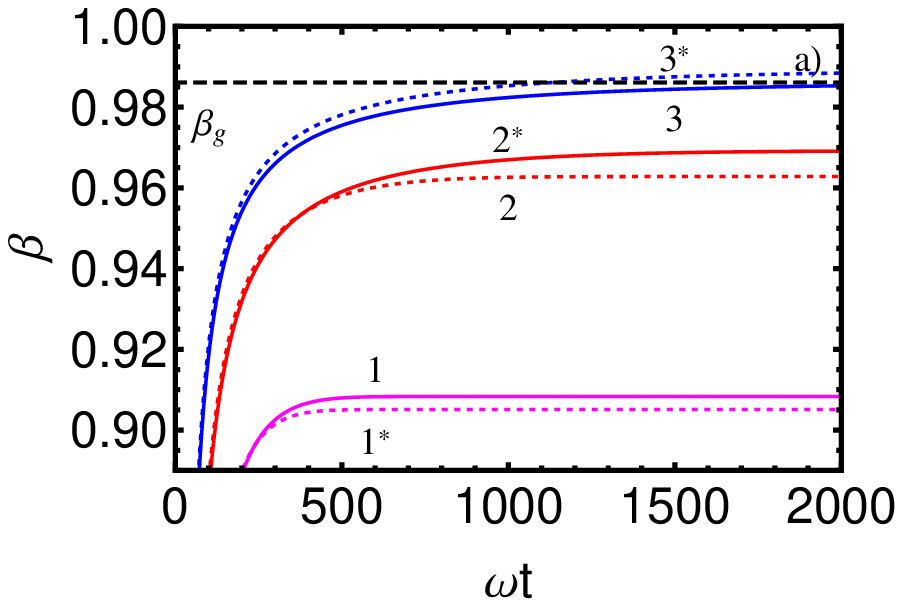} \epsfxsize5.5cm\epsffile{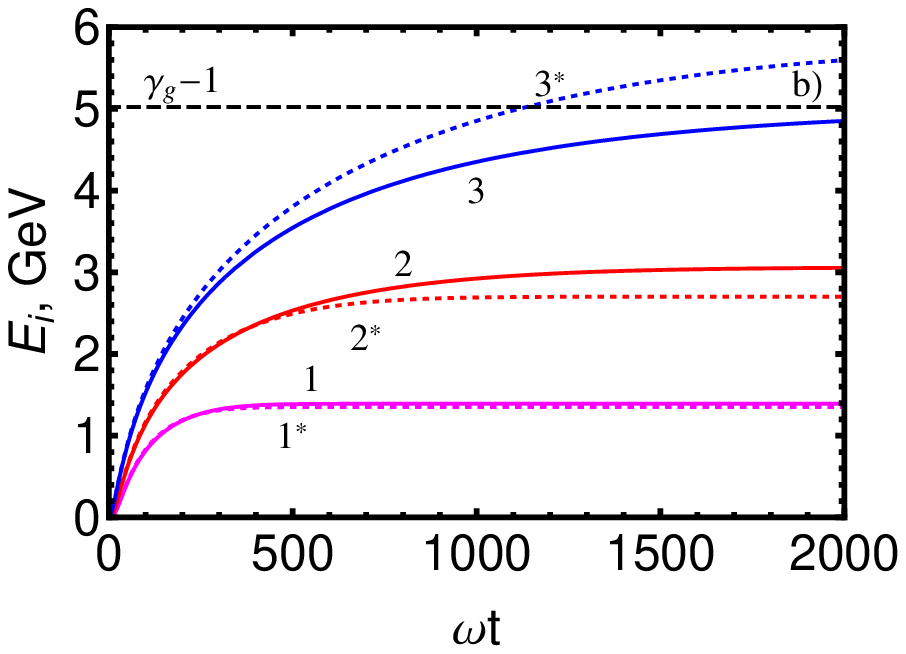} \epsfxsize6cm\epsffile{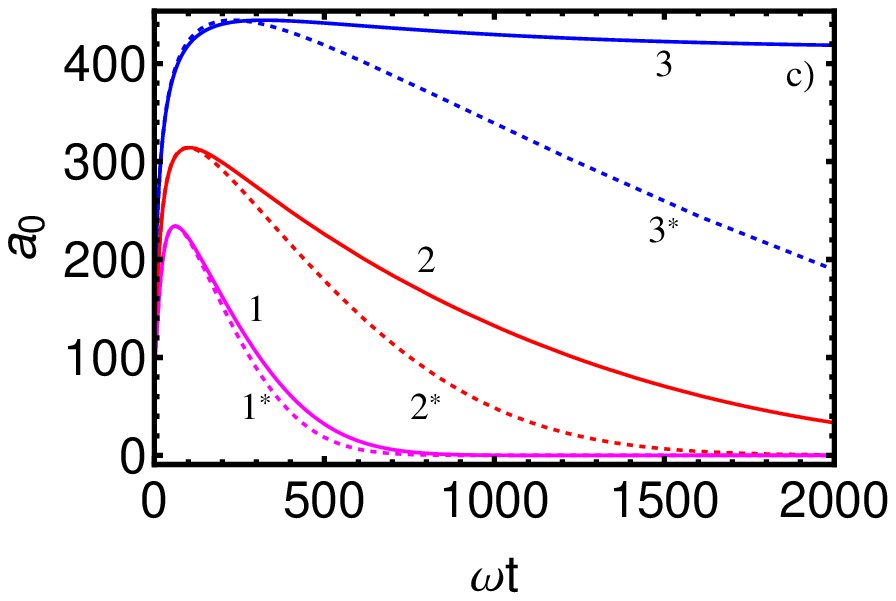} \epsfxsize5.5cm\epsffile{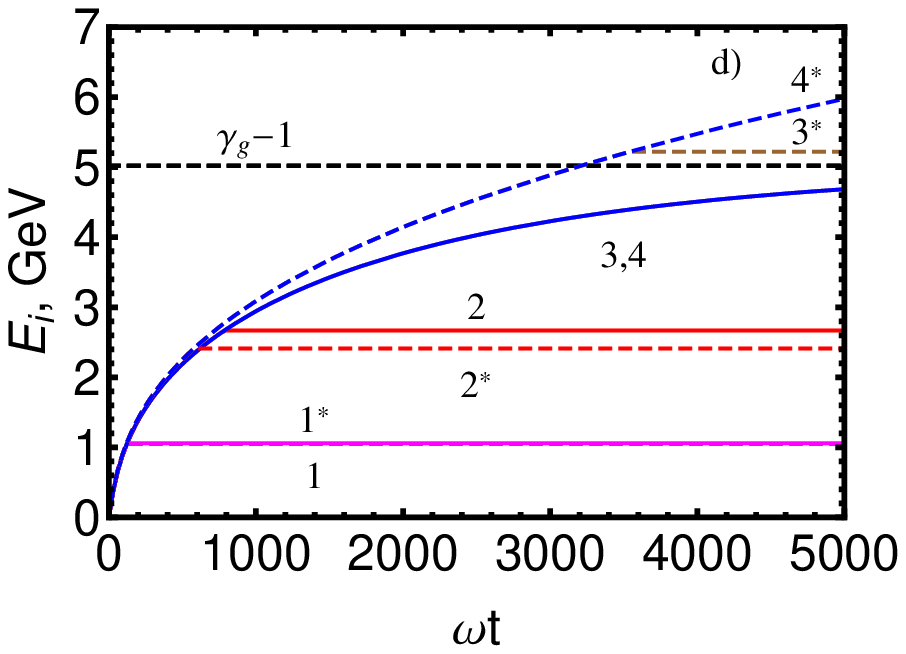}
\caption{(a) The dependences of the foil velocity on time for three values of the averaged laser pulse power: (1) 1 PW, (2) 1.8 PW, and (3) 3.6 PW for the cases of $\beta_g=0.986$ (solid curves) and $\beta_g=1$ (dashed curves), with the laser pulse duration 30 fs. The horizontal line marks $\beta_g$. (b) The dependence of the foil energy on time for three values of the laser pulse power: (1) 1 PW, (2) 1.8 PW, and (3) 3.6 PW for the cases of $\beta_g=0.986$ (solid curves) and $\beta_g=1$ (dashed curves), with the laser pulse duration 30 fs.  The horizontal line marks $\gamma_g-1$. (c) The evolution of the laser pulse EM field amplitude at the foil for three values of laser power: (1) 1 PW, (2) 1.8 PW, and (3) 3.6 PW for the cases of $\beta_g=0.986$ (solid curves) and $\beta_g=1$ (dashed curves). (d) The dependence of the foil energy on time for four values of the laser pulse duration: (1) 15 fs, (2) 30 fs, (3) 60 fs, and (4) 120 fs for the cases of $\beta_g<1$ (solid curves) and $\beta_g=1$ (dashed curves), with the laser pulse power 1.8 TW. The horizontal line marks $\gamma_g-1$. The asterisk mark the ($\beta_g=1$)-curves. The density of the foil is $n_e=400 n_{cr}$ and the thickness is $l=0.25\lambda$.}
\end{figure} 

From Eq. (\ref{K(beta)}) we can estimate the duration of the pulse for a given laser power for the pulse to become phase locked with the foil. We write down an equation for the phase corresponding to $\beta=\beta_g$:
\begin{equation}
K_\beta(\psi_{max})=\frac{1}{\beta_g^2(1-\beta_g^2)^{1/2}}\left(\frac{\pi}{2}-\arccos \beta_g\right),
\end{equation}
which for $K_\beta(\psi)=\kappa\psi$, i.e., for constant EM field, is reduced to 
\begin{equation}
\psi_{max}=\frac{1}{\kappa\beta_g^2(1-\beta_g^2)^{1/2}}\left(\frac{\pi}{2}-\arccos \beta_g\right),
\end{equation}
or for $\beta_g\rightarrow 1$, $\psi_{max}\approx (\pi/2)(\gamma_g/\kappa)$. The Eq. (21) indicates the minimal pulse duration for a given intensity that is needed to accelerate a foil up to the velocity equal to the group velocity of the laser pulse, which basically states a well known fact that, in the ultra relativistic case, almost all the laser energy is transferred to the ions \cite{RPA}. Moreover Eq. (\ref{eq of motion_phase}) is defined on the interval $0<\psi<\psi_{max}$. Since we can rewrite $\kappa$ as $\kappa=\gamma_{\beta_g=1}^{max}/\tau$, the maximum phase is 
\begin{equation}\label{max phase}
\psi_{max}=\frac{\pi}{2}\frac{\gamma_g}{\gamma_{\beta_g=1}^{max}}\tau.
\end{equation}
If $(\pi/2)\gamma_g>\gamma_{\beta_g=1}^{max}$, then $\psi_{max}>\tau$ and the foil can not be accelerated to $\gamma_g$, since there is not enough energy in the pulse. In this case the difference in maximum energy evolution in the cases of $\beta_g=1$ and $\beta_g<1$ should be small. For $\psi_{max}<\tau$ the difference should be rather large, since for $\beta_g<1$ the laser pulse would be able to accelerate ions only up to $\gamma_g-1$, while in the case of $\beta_g=1$ the ion maximum energy is determined by the total energy of the pulse. The dependence of the maximum ion energy evolution on the duration of the pulse for two cases of $\beta_g=1$ and $\beta_g<1$ is illustrated in Fig. 1d. Here the laser pulse is chosen to have a rectangular longitudinal profile. The curves corresponding to 15 fs laser pulse duration are indistinguishable for $\beta_g=1$ and $\beta_g<1$, which indicates that the group velocity effects are almost negligible. In the 30 fs case there is a slight difference, which is due to the dephasing of the laser pulse and the foil, and 60 fs and 120 fs cases demonstrate a significant difference. The $\beta_g<1$ curves are limited by $(\gamma_g-1)$, while $\beta_g=1$ curves go beyond this value. We can also see that $\beta_g<1$ curves for 60 fs and 120 fs are almost indistinguishable, since only the $\Delta\tau=\psi_{max}<\tau$ part of the pulse is able to accelerate the ions. There is no such similarity between the  60 fs and 120 fs cases, when $\beta_g=1$.

The limiting factor of $\beta_g<1$ can also be described in terms of the phase evolution. It is governed by the following equation 
\begin{equation}
\frac{d\psi}{dt}=\frac{\beta_g-\beta(\psi)}{\beta_g},
\end{equation}
which can be solved in quadratures:
\begin{equation}
t=\beta_g\gamma_g^2\left\{\beta_g\psi-\frac{\gamma_g}{\kappa\beta_g^{2}}\left[\ln\left(1-\frac{2}{1+\cot R+\csc R}\right)-\ln\left(-\frac{1}{\gamma_g\beta_g}+\left(1+\frac{1}{\gamma_g^{2}\beta_g^{2}}\right)^{1/2}\right)\right]\right\}.
\end{equation}
If $t\rightarrow \infty$, then depending on the ratio of $\gamma_{\beta_g=1}^{max}$ and $\gamma_g$ we have the following estimates for the phase evolution. (i) When $\gamma_{\beta_g=1}^{max}<\gamma_g$, the acceleration processes for $\beta_g=1$ and $\beta_g<1$ are indistinguishable, $R\ll\pi/2$:   
\begin{equation}
\psi=\frac{t}{\gamma_g^{2}\beta_g^{2}}.
\end{equation}
(ii) In the opposite case of $\gamma_{\beta_g=1}^{max}>\gamma_g$, the acceleration process is greatly affected by the existence of the fundamental limit on the maximum attainable ion energy, resulting in a different equation for the phase evolution:
\begin{equation}
\psi=\psi_{max}-\frac{2\gamma_g}{\kappa\beta_g^{2}}\exp\left(- \frac{\kappa\beta_g}{\gamma_g^{3}}t\right).
\end{equation}
Here we see that $\psi$ approaches $\psi_{max}$ but never exceeds it. Thus a laser pulse with duration $\tau>\psi_{max}$ will accelerate the foil up to $\beta=\beta_g$, but the part of the pulse characterized by the phase values $\psi_{max}<\psi<\tau$ will not be able to reach the foil and will not take a part in the acceleration process. 

\begin{figure}[tbp]
\epsfxsize6cm\epsffile{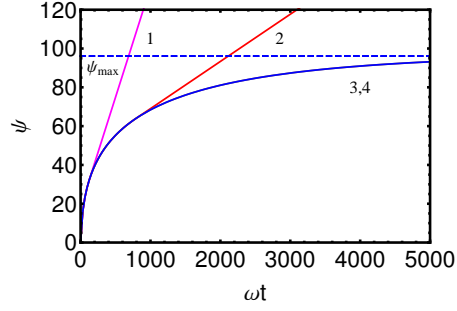} 
\caption{The dependence of the phase $\psi$ on time for a 1 PW, f/D=1.5 laser pulse for four durations of the pulse: 13 fs (1), 27 fs (2), 54 fs (3), and 108 fs (4). The maximum phase from Eq. (\ref{max phase}) is plotted as a dashed horizontal line. The density of the foil is $n_e=400 n_{cr}$ and the thickness is $l=0.25\lambda$.}
\end{figure}

In Fig. 2 we demonstrate the dependence of the phase on time. For $\gamma_{\beta_g=1}^{max}<\gamma_g$ (curves 1 and 2) the phase tends to infinity as the time tends to infinity, which means that the pulse has accelerated the foil to some energy and is no longer interacting with the foil. In the case $\gamma_{\beta_g=1}^{max}>\gamma_g$ (curves 3 and 4) the phase tends to $\psi_{max}$, i.e., the pulse becomes phase locked with the foil. When phase locked, accceleration has stopped, since the laser is no longer able to transfer energy to the foil. Thus the laser pulse and the foil co-propagate without interaction and $\psi=\psi_{max}$ corresponds to the foil reaching the velocity equal to the group velocity of the pulse.    

Thus, following the approach of Ref. \cite{slow wave_PRL} we showed that there exists a fundamental limit on the maximum attainable ion energy in the RPA regime and it is equal to the $\gamma$-factor of the driving laser pulse. This limit greatly affects the acceleration process for the pulses with $\gamma_{\beta_g=1}^{max}>\gamma_g$. We also identified a minimum laser duration needed to reach the maximum accelerated ion energy for fixed laser intensity, $I_L$: 
\begin{equation}
\tau_{min}=\frac{\pi}{2}\frac{\gamma_g n_e l}{I_L}.
\end{equation}  

\section{Transverse expansion of the target}

A group velocity smaller than the vacuum light speed appears naturally in the case 
of tightly focused laser pulses \cite{Review}, which diverge rather quickly after passing 
through the focus. Assuming that this divergence forces the irradiated part of the foil to expand, 
following the increase of the laser spot size, we study how the transverse expansion of the target 
limits the maximum attainable ion energy during the RPA, and whether this limitation dominates over the fundamental 
effects of the group velocity. Assuming that the field of the pulse can be given by the paraxial approximation, 
characterized by the laser pulse waist at focus, $w_0$ and the Rayleigh length $L_R=\pi w_0^2/\lambda$, 
the evolution of the laser pulse waist as it travels away from focus is $w(x)=w_0\left[1+(x/L_R)^2\right]^{1/2}$,  
the amplitude of the field scales with the distance from the focus as $a(x)=a_0\left[1+(x/L_R)^2\right]^{-1/2}$, 
and the group velocity is $\beta_g\simeq 1-1/k^2w_0^2$ \cite{group velocity}. Since we are interested 
in the maximum ion energy, we consider RPA of an on-axis element of the foil. The intensity profile 
near the axis can be approximated by an expanding spherical cup with curvature radius equal 
to the laser waist, $w(x)$. The on-axis element of the foil can also be approximated by an expanding 
spherical cup with the curvature $w(x)$ and areal density  equal to $n_e l=n_0l_0\left[1+(x/L_R)^2\right]^{-1}$ 
and $\varepsilon_e(x) = \varepsilon_e(0)\left[1+(x/L_R)^2\right]^{-1}$. Substituting 
the field and areal density into Eq. (\ref{eq of motion}), we see that the right hand side 
of Eq. (\ref{eq of motion}) depends on the distance from the focus only through the reflection 
coefficient \cite{optimized RPA}: 
\begin{equation}\label{eq of motion_x}
\frac{d\beta}{dt}=|\rho(x)|^2\frac{m_e}{m_i}\frac{a^2(\psi)}{\varepsilon_e}\beta_g(1-\beta^2)^{1/2}(\beta_g-\beta)(1-\beta\beta_g),
\end{equation} 
where   
\begin{equation}
\rho(x)=\frac{\tilde{\epsilon}_e(x)}{a(x)}\left[\frac{b(x)}{(1+b(x)^2)^{1/2}}\right],
\end{equation} 
and
\begin{equation}
b(x)=\frac{1}{2^{1/2}}\left\{\left[(a(x)^2-\tilde{\epsilon}_e^2(x)-1)^2+4a^2(x)\right]^{1/2}+(a^2(x)-\tilde{\epsilon}_e^2(x)-1)\right\}^{1/2}.
\end{equation} 
In what follows we solve Eq. (\ref{eq of motion_x}) numerically, taking into account transverse expansion of the foil and laser pulse divergence. The comparison between the cases with the transverse expansion taken into account and without is shown in Fig. 3 for a $1.8$ PW laser pulse interacting with a $0.25\lambda$ thick hydrogen foil with the density of $n_e=400 n_{cr}$ for three values of the f-number: $f/D=1$, $f/D=1.5$, and $f/D=2$. In order to demonstrate the effect of transverse expansion and laser divergence we show in Fig. 3a two sets of curves: one corresponding to f/D=1, the other to f/D=1.5. Each set consists of three curves: two curves with no target expansion and laser divergence, with $\beta_g=1$ and $\beta_g<1$, and the third one with transverse expansion and $\beta_g<1$. As one can see this effect significantly reduces the maximum ion energy by switching off the acceleration early. This switching off can clearly be seen in Fig. 3b, where the evolution of the reflected EM field amplitude at the foil is given for each of the curves in Fig. 1a. If the transverse expansion is taken into account, the EM field amplitude is much lower, which has a direct consequence in lower maximum ion energy.  

\begin{figure}[tbp]
\epsfxsize6cm\epsffile{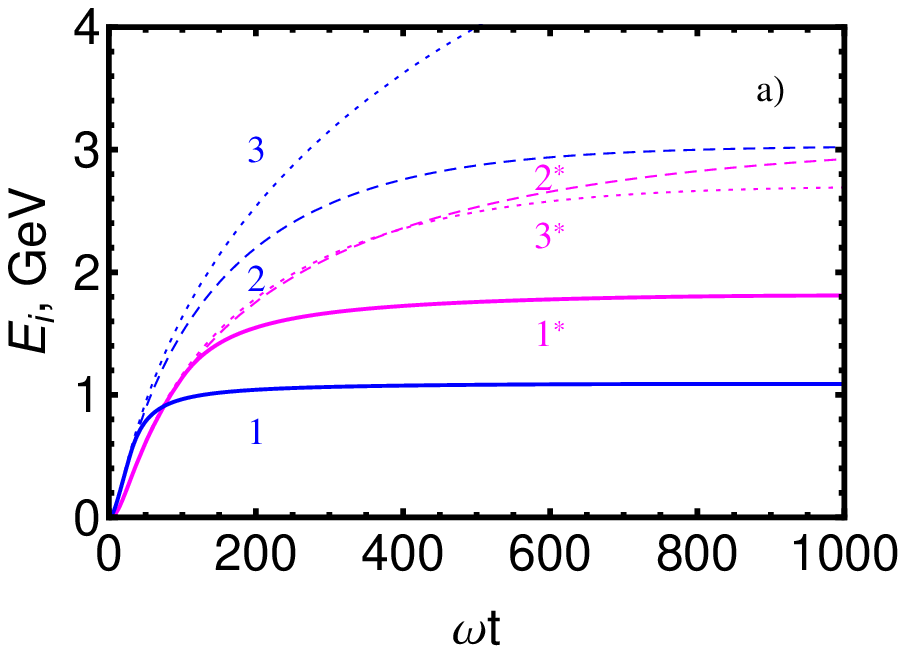} \epsfxsize6.5cm\epsffile{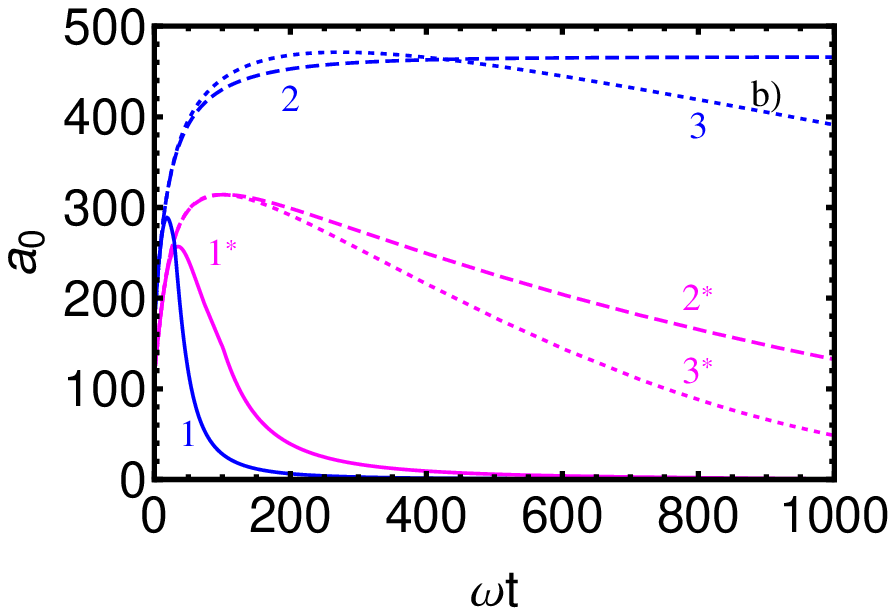} \epsfxsize6cm\epsffile{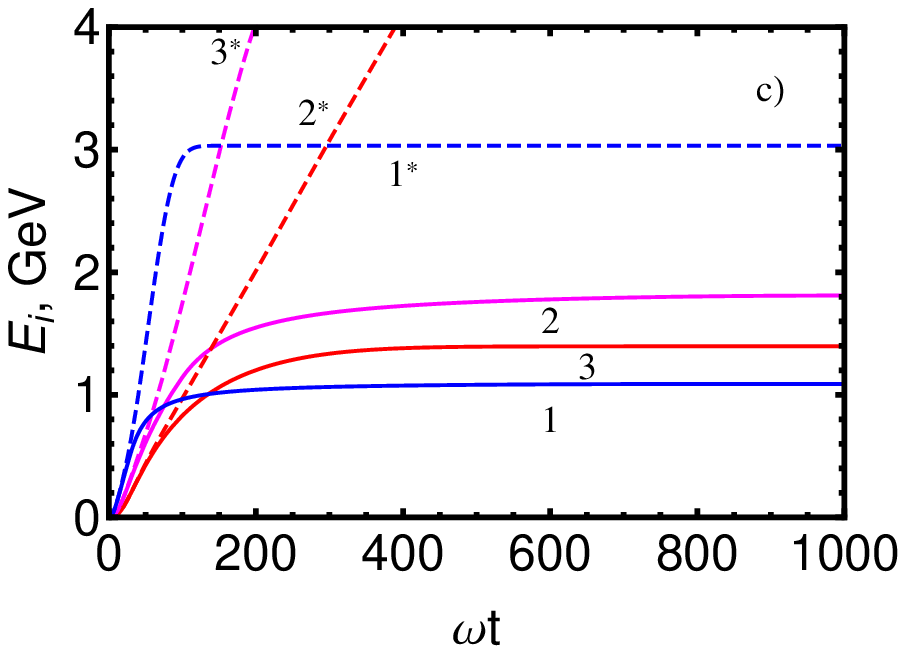} \epsfxsize6.5cm\epsffile{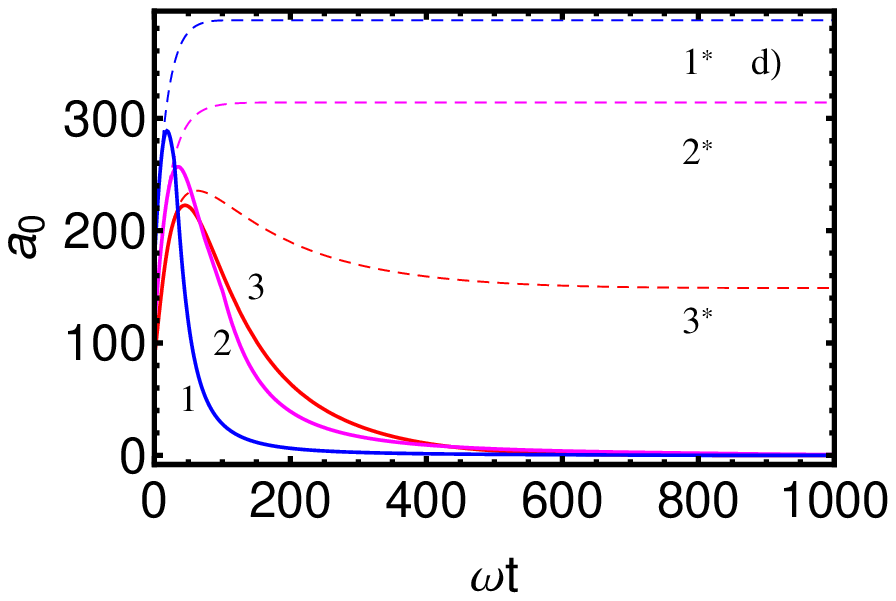}
\caption{(a) The evolution of the maximum ion energy for two values of f-number: f/D=1 (blue curves, 1, 2, and 3) and f/D=1.5 (magenta curves, 1$^*$, 2$^*$, and 3$^*$). The solid curves (1 and 1$^*$) are the solutions of Eq. (\ref{eq of motion_x}) with laser divergence and target expansion taken into account. The dotted curves (2 and 2$^*$) are the solutions of Eq. (\ref{eq of motion}) with $\beta_g=1$ and dashed curves (3 and 3$^*$) are the solutions of Eq. (\ref{eq of motion}) with $\beta_g<1$. (b) The evolution of EM field amplitude at the foil corresponding to the curves in Fig. 3a. (c) The evolution of the maximum ion energy for f/D=1 (blue curves, 1 and 1$^*$), f/D=1.5 (magenta curves, 2 and 2$^*$), and f/D=2 (red curves, 3 and 3$^*$). The curves 1, 2, and 3 are the solutions of Eq. (\ref{eq of motion_x}), while the curves 1$^*$, 2$^*$, and 3$^*$ correspond to the guided case, where the laser is not diffracting and the reflection coefficient is equal to one. (d) The evolution of EM field amplitude at the foil corresponding to the curves in Fig. 3c. Laser pulse power is 1.8 PW, duration is 30 fs, the foil thickness is $0.25\lambda$, and density is $n_e=400 n_{cr}$.}
\end{figure}

The utilization of an external guiding structure may relax the limits on maximum attainable ion energy. To model such interaction we solve Eq. (\ref{eq of motion}), assuming that the laser pulse is guided by a self-generated channel with a transverse size of $w_0=0.9\lambda$ ($f/D=1$), $w_0=1.35$ ($f/D=1.5$), and $w_0=1.8$ ($f/D=2$), and the foil stays opaque for the pulse ($|\rho|=1$). The solutions for such configuration are shown by dashed curves in Fig. 3c. One can see that the external guiding structure significantly enhances the maximum attainable ion energy, which is now limited by the laser group velocity in such a structure. This limit is illustrated in Fig. 3d, where the evolution of the reflected EM field amplitude is shown. The fields corresponding to the cases with guiding demonstrate the phase locking with the ions, \textit{i.e.}, they co-propagate with the foil, but are not able to transfer any energy to it. In principle, a composite target, consisting of a thin foil, followed by a near critical density (NCD) slab, may provide an example of such guiding. The laser pulse will accelerate the irradiated part of the foil in the self-generated channel in the NCD plasma \cite{MVA}. Though the foil density will drop due to the transverse expansion, the NCD plasma electrons being snowplowed by the pulse would provide an opaque density spike, which being pushed by the radiation pressure would drag the ions of the foil with it. Thus such configuration is similar to the one considered above: the laser pulse is guided with no diffraction and, although the density of ion decreases, the reflection coefficient is unity.          
 
\section{Composite target RPA: Results of 2D PIC simulations} 

\begin{figure}[tbp]
\epsfxsize7cm\epsffile{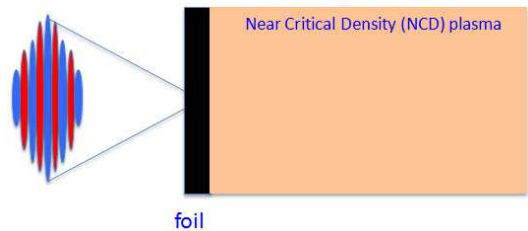}
\caption{The simulation setup: laser is focused on a thin foil, which is followed by a near critical density plasma.}\label{setup}
\end{figure}

In this section we present the results of 2D PIC simulations (using the code REMP \cite{REMP}), of an intense laser  interacting with two types of targets: (i) a single thin solid density foil and (ii) a thin solid density foil followed by an NCD plasma slab (a composite target). As was shown in the previous section a composite target should provide guiding for the laser pulse and compensate for the transverse expansion of the foil, leading to higher proton energies than in the case of a single foil. The simulation box is $100\times60\lambda^2$, $dx=dy=0.025\lambda$, $dt=0.0125 \times 2\pi/\omega$, and the number of particles per cell is 100. The Gaussian laser pulse is initialized at the left border with dimensionless potential $a_0=100$, waist $w=4\lambda$ and duration $\tau=30$ fs, which corresponds to the average power of 1.8 PW. The pulse is focused with f-number $f/D=2$ at the left front size of the target, which is placed 16$\lambda$ away from the left border. The composite target consists of a fully ionized hydrogen foil and a hydrogen NCD plasma slab placed right behind the foil. The typical simulation setup is shown in Fig. \ref{setup}.

\begin{figure}[tbp]
\epsfxsize15cm\epsffile{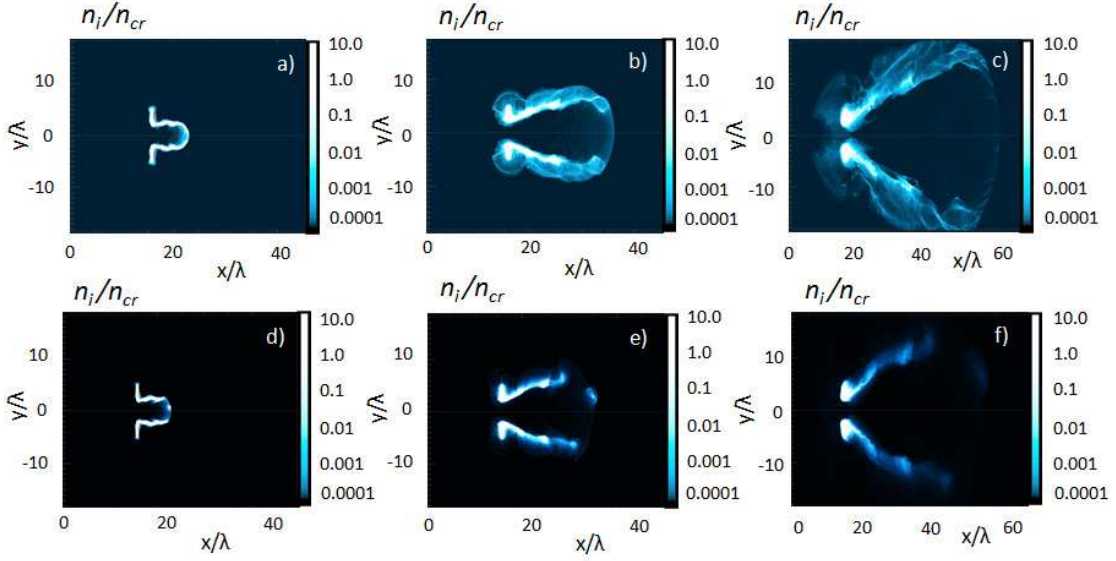}
\caption{The evolution of the density of the ions originating from the foil during the laser pulse interaction with a composite target (a-c) and a single foil (d-f) at t=35 (a,d), t=50 (b,e), t=75 (c,f). } \label{ion density}
\end{figure}

\begin{figure}[tbp] 
\epsfxsize15cm\epsffile{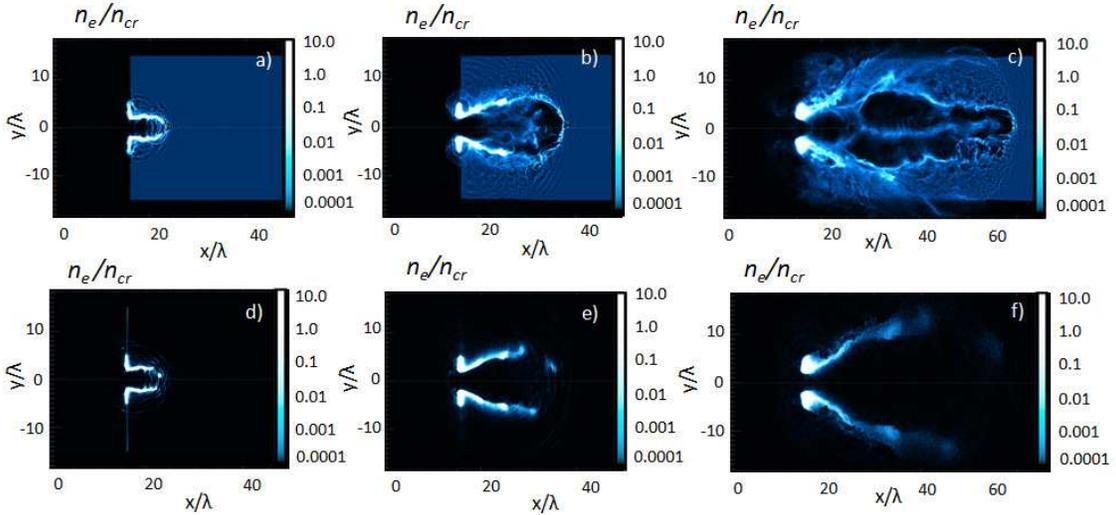}
\caption{The evolution of the density of the electrons during the laser pulse interaction with a composite target (a-c) and a single foil (d-f) at t=35 (a,d), t=50 (b,e), t=75 (c,f). }\label{electron density}
\end{figure}

In Fig. \ref{ion density} we present the ion density distributions for the cases of a composite target and a single foil at different moments in time. The laser peak power is 1.8 PW and is focused with f-number=2 at the foil. The foil is $0.25\lambda$ thick and the NCD plasma slab is $50\lambda$ thick and its density is equal to critical one, $n_e=n_{cr}$. In both cases the transverse expansion of the target plays an important role, leading to a significant deformation of the foil and density redistribution. As a result the density near the laser axis is reduced. In the case of a single foil it means that the foil became transparent for radiation and the acceleration has stopped. In the case of a composite target it is not so, since the background electrons from the NCD plasma being snowplowed by the laser compensate for the density decrease. Such electron behavior is shown in Fig. \ref{electron density}. In the case of a single foil the electrons are subjected to transverse expansion, which is similar to that shown in Fig. \ref{ion density} for the ion density distribution. In the case of a composite target, the electrons that were originally in the foil are also pushed to the side but the NCD electrons being snowplowed by the laser pulse form a dense shell in front of the laser, providing a robust relativistic mirror for proton acceleration up to the end of the NCD plasma slab. The different behavior of the electron component of plasma in the cases of a single foil and a composite target should translate into different values of the maximum ion energy.   

\begin{figure}[tbp]
\begin{center}
\epsfxsize7cm\epsffile{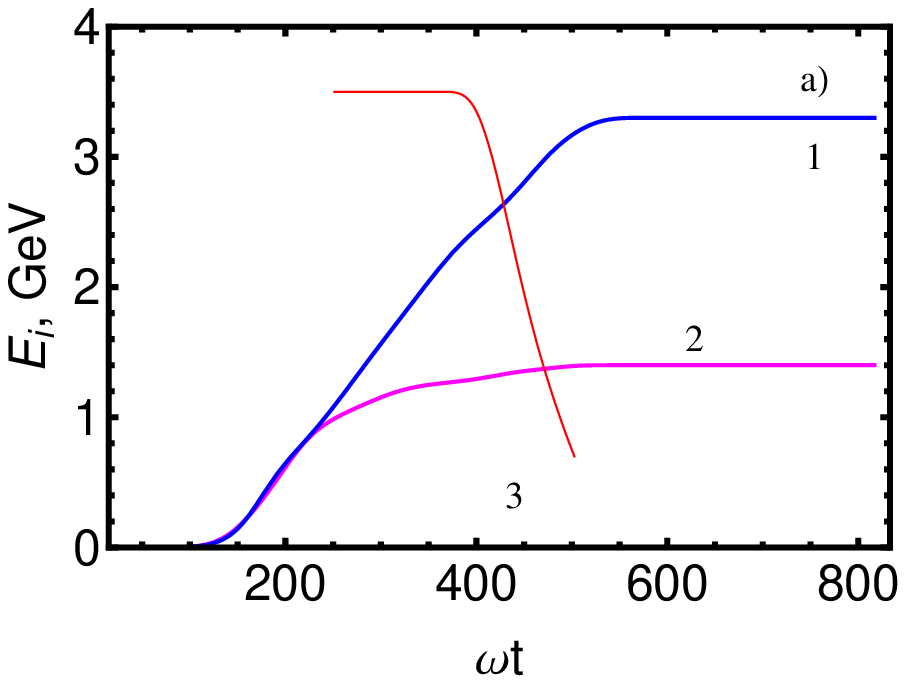}
\epsfxsize7cm\epsffile{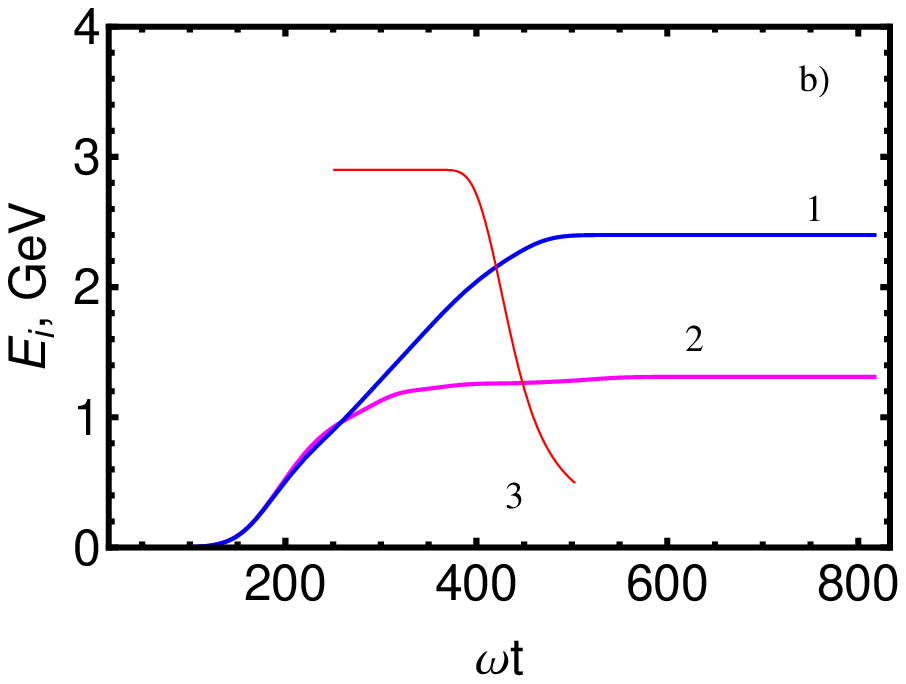}
\end{center}
\caption{The time evolution of the ion energy for a composite target (blue curves, 1) 
RPA and single foil (magenta curves, 2) RPA. The foil thickness is $0.25\lambda$ 
with densities $n_e=225 n_{cr}$ (a) and $n_e=400 n_{cr}$ (b). 
The thickness of the NCD plasma slab is $50 \lambda$ and density is equal to $n_{cr}$. 
The evolution of $\gamma_I-1$ is shown by thin red curves (3).} \label{ion energy evolution}
\end{figure}

The evolution of the maximum ion energy in both cases is shown in Fig. \ref{ion energy evolution} for two densities of the foil: $n_e=225 n_{cr}$ (Fig. \ref{ion energy evolution}a) and $n_e=400 n_{cr}$ (Fig. \ref{ion energy evolution}b). One can see that in the case of a composite target the ion maximum energy is significantly higher, which is due to the longer acceleration length in the NCD plasma. In the case of a single foil, the target becomes transparent at $t\approx 35$, when, according to Fig. 5, the density on-axis drops several times below the initial one. The final energy for the single foil is in good agreement with the analytical results of the previous section, shown in Fig. 3. The maximum ion energy in the case of a composite target should be determined by the group velocity of the laser in the self-generated plasma channel. However, due to the fast depletion of the pulse as well as its reflection at laser-plasma interface, the laser group velocity can not unambiguously be determined from the results of PIC simulations. In this case, we chose the velocity of the of the laser-plasma interface, $\beta_I$, as the characteristic quantity for ion acceleration in the channel.
Using results of Ref. \cite{ShFr} we can find the normalized velocity laser-plasma interface from the condition 
that it takes the depletion energy time, $t_{depl}$, for the laser pulse tail to reach the laser-plasma interface.
This yields  
\begin{equation}
\beta_I=\beta_g\left(1-\frac{n_e }{n_{cr} a}\right)\approx \beta_g\left(1-\frac{1}{\gamma_g^2}\right)=\beta_g^3,
\end{equation}
i.e., 
\begin{equation}
\gamma_I\approx\frac{\gamma_g}{3^{1/2}}
\end{equation} 
(for the definition of the group velocity of the laser pulse inside the self-generated channel in the NCD plasma, $\gamma_g$, see Appendix A). For the case considered $\gamma_I\approx 5$, which is in good agreement with the results of PIC simulations, taking into account pulse reflection during the initial interaction with the foil. In Fig. \ref{ion energy evolution}, the energy $\gamma_I-1$ (from PIC results) is shown by thin red curves. Note that $\gamma_I-1$ remains constant during the pulse propagation through the NCD plasma, but after the pulse depletion becomes significant, it decreases, marking the end of the ion acceleration for both values of the foil density studied in simulations. We note that the ion energy is not able to reach the maximum value of $\gamma_I-1$ due to the laser pulse depletion in NCD plasma. 

\begin{figure}[tbp]
\begin{center}
\epsfxsize7cm\epsffile{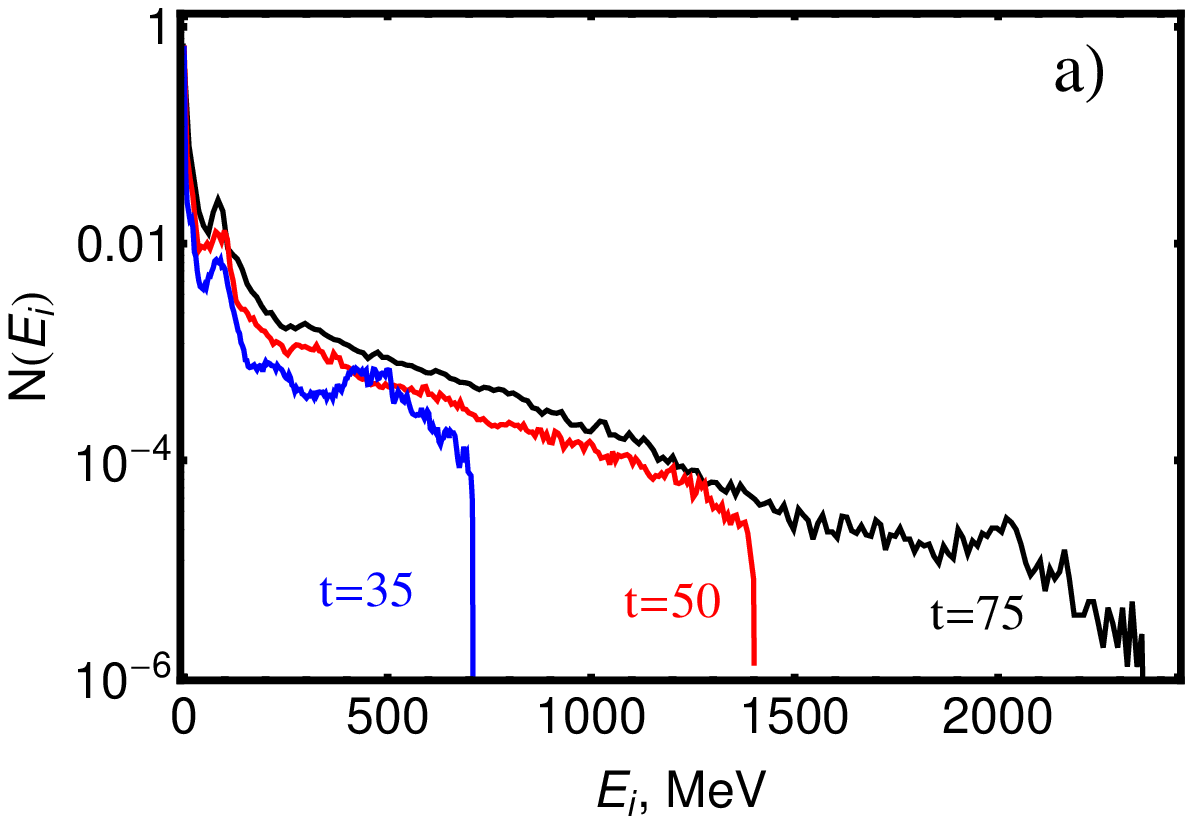}
\epsfxsize7cm\epsffile{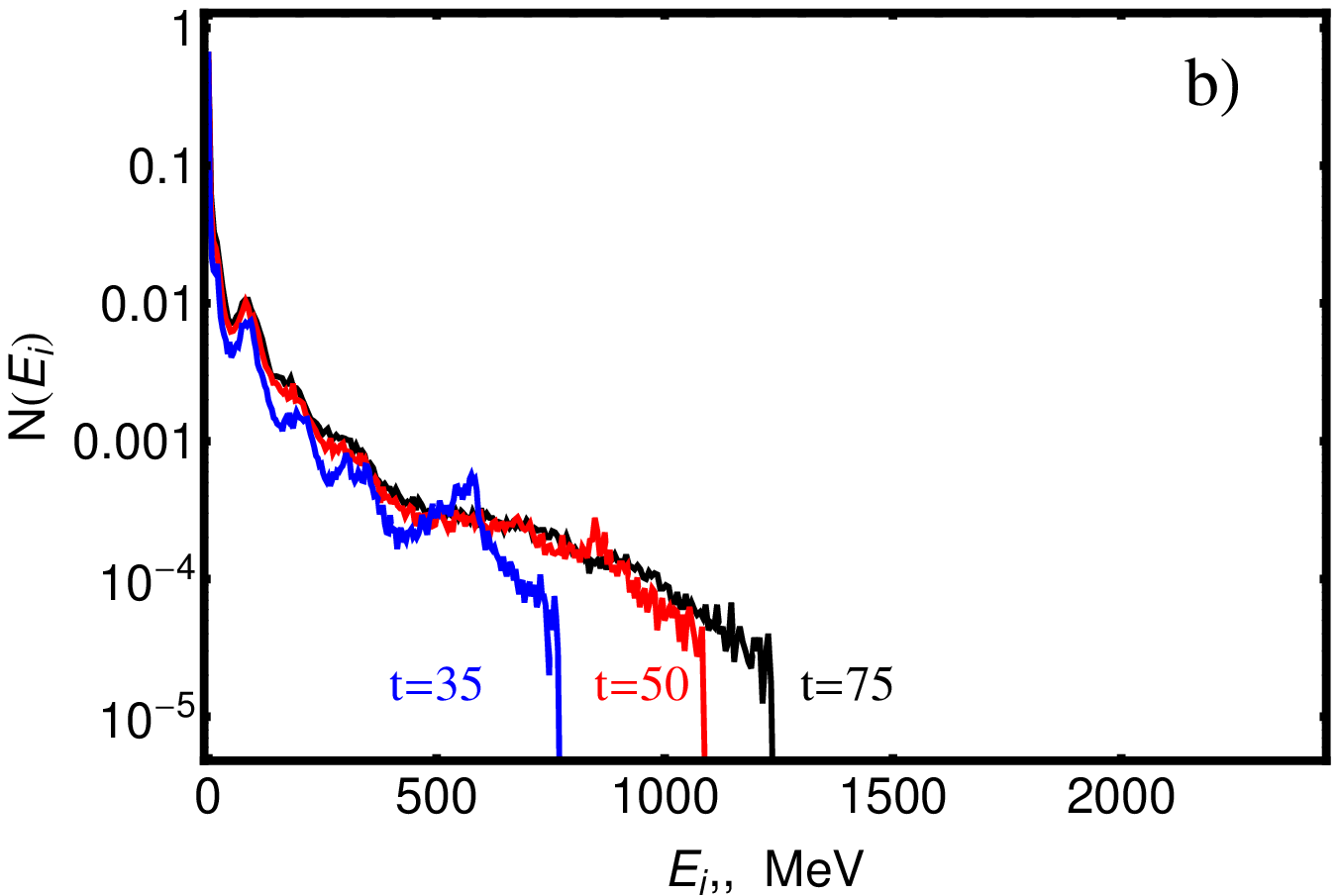}
\end{center}
\caption{The evolution of the ion spectrum originating from the foil during the laser pulse interaction 
with a composite target (a) and a single foil (b) at  t=35 (red curve), t=50 (blue curve), and t=75 (black curve).  
The parameters of the interaction are the same as in Fig. 3.}\label{spectrum of ions}
\end{figure}

In order to further illustrate the differences between these two cases, we show in Fig. \ref{spectrum of ions} the spectra of protons in the case of a composite target (Fig. \ref{spectrum of ions}a) and a single foil (Fig. \ref{spectrum of ions}b) at different moments in time. While at $t=35$ the spectra look similar, at $t=50$ the difference becomes pronounced, since a single foil became transparent for radiation, marking the termination of the acceleration process. Moreover, at $t=75$, in the case of a composite target, the protons gained almost 50\% of their energy at $t=50$, and, in the case of a single foil, the gain is as low as 10\%. 

\begin{figure}[tbp]
\epsfxsize7cm\epsffile{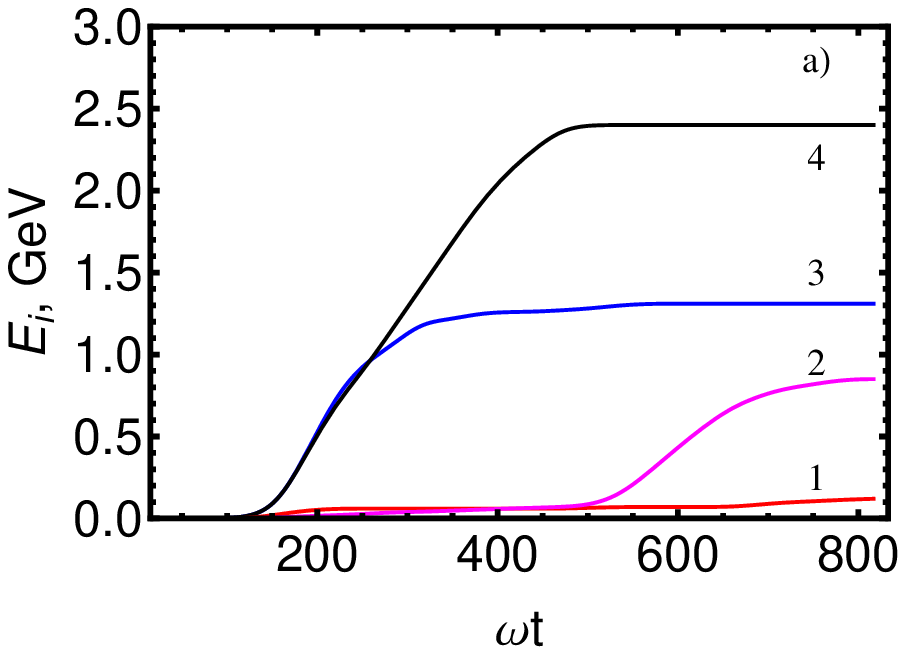}\epsfxsize7cm\epsffile{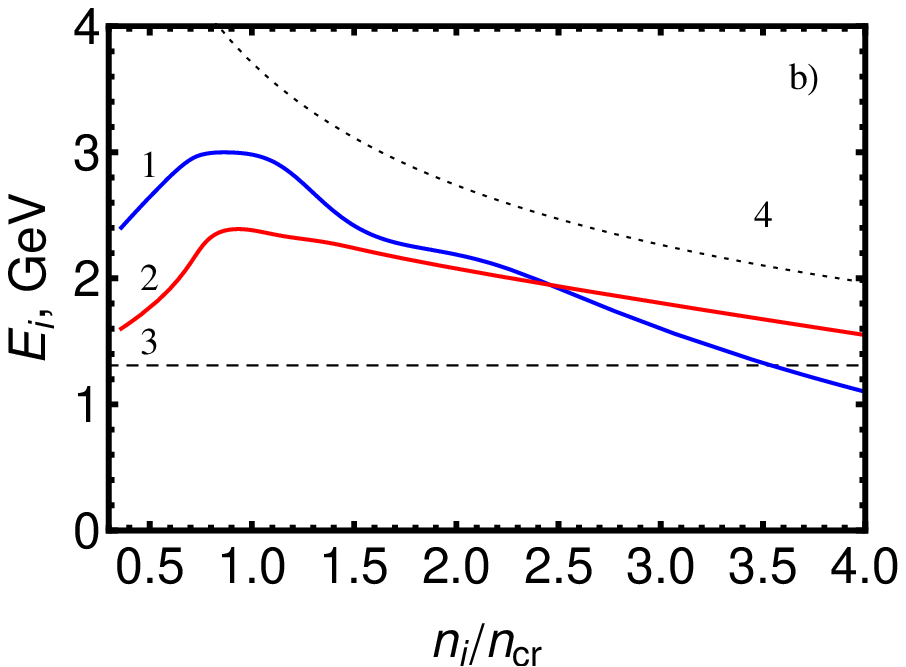}  
\caption{a) The evolution of the maximum ion energy in the case of (1) the protons from the NCD plasma slab (composite target), (2) the protons from the NCD plasma slab (no foil attached), (3) the protons from the foil (single foil target), and (4) the protons from the foil (composite target); b) The dependence of the maximum ion energy on the density of the NCD plasma slab for two values of the foil density $n_e=225 n_{cr}$ (blue curve , 1) and $n_e=400 n_{cr}$ (red curve, 2). The dashed curve (3) marks the maximum proton energy in the case of a single foil. The dotted curve is the dependence of $\gamma_I-1$ on density.}\label{comparison}
\end{figure}

In Fig. \ref{comparison}a we show that the protons can originate from different parts of a composite target. If we take such target apart, each of its parts, when irradiated by an intense laser, can generate high energy protons. We compared the  maximum energy evolutions in all these cases to find that the protons from the foil, in the case of a composite target, acquire maximum energy of 2.4 GeV. Note that the energy of protons from a single NCD slab is in good agreement with the results of the analytical estimate (see Appendix B). Thus, the addition of an NCD plasma slab at the back of a thin foil, is a crucial feature of the target design. 

In Fig. \ref{comparison}b we show the dependence of the maximum ion energy on the density of the NCD plasma. Since the acceleration is characterized by the velocity of the laser-plasma interface, it is plausible to assume that the energy of ions would scale as $\gamma_I-1$ scales with density. One can see from Fig. 8b that maximum ion energy has similar dependence on density as $\gamma_I-1$. Here we took into account the fact that a significant part ($\sim50\%$) of the laser pulse was reflected at the foil before the remaining part of the laser pushed the irradiated part of the foil inside the NCD plasma. However, for low densities, the similarity between $\gamma_I-1$ and the maximum proton energy fails, due to the fact that the laser propagation through such plasma can no longer be described as a propagation inside a self-generated channel \cite{MVA}.     

Thus we identified the dependence of the maximum ion energy on the density of the NCD plasma and the fact that this energy is characterized by the velocity of the laser-plasma interface. We compared the cases of a single foil and a composite target, showing that in both cases the transverse expansion significantly modifies the target. In the case of a single foil, this modification leads to the termination of the acceleration due to the increased transparency of the target. In the case of a composite target the NCD electrons being snowplowed by the laser pulse form a dense shell in front of the laser, providing a robust relativistic mirror for proton acceleration.        

\section{Conclusions}

{\noindent}We considered laser driven acceleration of ions in the radiation pressure acceleration (RPA) regime, taking into account the fact that the laser group velocity can be smaller than the light vacuum speed. We also included the deformation of the thin foil target, i.e., the transverse expansion in the equations of motion of the foil. This allowed us to identify two factors that limit the maximum attainable in the RPA ion energy: (i) the fundamental effect of the laser pulse group velocity being less than vacuum light speed, and (ii) the transverse expansion of the target, which plays a major role when tightly focused pulses are used. The first one results in the energy transfer from the pulse to the foil vanishing as the velocity of the foil approaches the group velocity of the laser. Thus the ion velocity can not exceed the laser group velocity. The transverse expansion of the target effectively decreases the density of the foil, eventually making it transparent for radiation. For realistic parameters of the laser foil interaction the effect of the transverse expansion dominates over the group velocity effect. 

We showed that the utilization of an external guiding may relax the constraints on maximum attainable ion energy. Such guiding should accommodate the transverse expansion of the ion component without the foil becoming transparent. We showed that a composite target, a thin foil followed by an NCD slab, is an example of such guiding structure. The NCD slab provided guiding of the laser pulse during the acceleration process as well as the electrons to replace the ones that are expelled due to the transverse expansion. The comparison of a single foil RPA and a composite target RPA shows that in the latter case the ions can gain the energy several times larger than in the former case, thus greatly increasing the effectiveness of the RPA regime of laser driven ion acceleration. In such a configuration the group velocity effects begin to dominate and determine the maximum achievable ion energy.     

\acknowledgements

We acknowledge support from the NSF under Grant No. PHY-0935197 and the Office of Science of the US DOE under Contract No. DE-AC02-05CH11231 and No. DE-FG02-12ER41798 and the Ministry of Education, Youth and Sports of the Czech Republic (ELI-Beamlines reg. No. CZ.1.05/1.1.00/02.0061). The authors would like to thank for discussions C. Benedetti, M. Chen, C. G. R. Geddes, L. Yu, D. Margarone, and G. Korn.  

\appendix
\section{Laser propagation in Near Critical Density Plasma}

Here we address the propagation of an intense laser pulse in a near critical density plasma. The proposed composite target concept depends on the self-channeling of an intense laser pulse in such a plasma. The determination of the channel parameters, radius and length, the parameters of the laser pulse mode, which will propagate in this channel, are prerequisite for the study of the proposed mechanism. In principle, the propagation of an intense laser pulse in the relativistically underdense plasma can be approximated by the propagation of an EM wave in a waveguide \cite{MVA}. It is connected with the fact that, as the laser pulse propagates in such a plasma, the ponderomotive force of the laser expels electrons and ions from the region of the laser pulse. However due to much smaller charge to mass ratio the time of ion response is much larger than that of electrons, which leads to the formation of a positively charged region just after the front of the laser pulse. The expelled electrons are subject to the laser ponderomotive force and the attracting charge separation fields. This results in the formation of an electron density channel with high density walls. After the laser pulse passes, the ions also expand in the transverse direction forming an ion density channel. This is partly due to the response to the ponderomotive force and partly due to the coulomb explosion of a positively charged channel background. The laser radiation at the early stages of channel evolution is not able to penetrate these walls and is totally contained inside the channel. Then, the laser mode with lowest transverse frequency, an H-wave ($E_x=0$), will propagate through this channel:
\begin{equation}
H_x=A J_1(\Omega r)\cos\left(\omega t - kx \right),~~~E_r=\frac{i\omega}{\Omega^2}\frac{\partial H_x}{\partial r},
~~~H_r=\frac{\omega}{\Omega^2}\frac{\partial H_x}{\partial r},
\end{equation}
where $J_1$ is the Bessel function of the first kind and $\Omega=1.84/R$; $R$ is the radius of the channel, and $A=(2\Omega/\omega)(m\omega a/e)$, here $a$ is defined as the amplitude of the transverse electric field. The radius of the channel can be estimated from balancing the electron energy gains from the laser pulse EM field and from the field generated by the positive charge of ions in the channel: $R_{ch}=\sqrt{a}\left(\omega/\omega_p\right)\lambda/\pi$. The group velocity of such an EM wave in the waveguide is $\beta_g=\left[1-(1.84/2)^2(\omega_p/\omega)^2a^{-1}\right]$. The total energy of such a pulse in the channel is
\begin{equation}
\mathcal{E}_L^{ch}=\pi R^2 \tau a^2 m_e n_{cr} K,
\end{equation}
where $\tau$ is the pulse duration and the factor $K$ comes from the integration over transverse coordinates and the duration of the laser pulse 
\begin{equation}
K=\sqrt{\frac{\pi}{32}}\left[J_1(\kappa R)^2-J_0(\kappa R)J_2(\kappa R)\right]\approx\frac{1}{13.5}.
\end{equation}
Here $J_0$ and $J_2$ are Bessel functions of the zero and second order respectively. The relation between the amplitude of the laser pulse vector-potential and the laser pulse power, $P=\mathcal{E}_L^{ch}/\tau$, is
\begin{equation}\label{a_ch}
a_{ch}=\left(\frac{2}{K}\frac{P}{P_c}\frac{n_e}{n_{cr}}\right)^{1/3},
\end{equation}  
where $P_c=2 m_e^2/e^2=17$ GW. For $P=1$ PW and $n_e/n_{cr}=1$, the relation yields $a\approx 116$. As was shown in Ref. \cite{MVA} the depletion length of the laser pulse in the NCD plasma is $L_{ch}=a (n_{cr}/n_e)L_p K$, where $L_p$ is the length of the laser pulse. Using $R_{ch}=\sqrt{a}\left(\omega/\omega_p\right)\lambda/\pi$ we can write the depletion length, normalized to pulse length, and the radius of the channel, normalized to laser wavelength, in terms of the laser pulse power:
\begin{equation} \label{depletion length}
\frac{L_{ch}}{L_p}=2^{1/3} K^{2/3}\left(\frac{n_{cr}}{n_e}\right)^{2/3}\frac{P}{P_c},
\end{equation}
\begin{equation} \label{channel radius}
\frac{R_{ch}}{\lambda}=\frac{1}{\pi}\left(\frac{n_{cr}}{n_e}\right)^{1/3}\left(\frac{2P}{K P_c}\right)^{1/6}.
\end{equation}
The $\gamma$-factor, corresponding to the group velocity of the EM wave in the waveguide, neglecting pulse depletion and evolution, is 
\begin{equation}\label{gamma_g^ch}
\gamma_g^{ch}\sim\left(\frac{2 P}{KP_c}\right)^{1/6}\left(\frac{n_{cr}}{n_{e}}\right)^{1/3}.
\end{equation}
If a thin foil is accelerated by the radiation pressure of an EM pulse inside this channel, then $\gamma_g^{ch}$ is the maximum achievable energy of the ions.

\section{Maximum ion energy via MVA mechanism}

In what follows we estimate the maximum ion energy that can be gained via the MVA mechanism in an NCD plasma \cite{helium}. In the MVA regime, the acceleration is achieved by an electric field at the back of the target, which is generated by, and is of the order of, the magnetic field ($B_{ch}$) of electrons, accelerated by the laser pulse in plasma in the forward direction, as they exit the target. This EM field, associated with the electron current leaving the target expands along the back surface of the target, while the magnetic field flux is conserved. 
The maximum energy gain of a charged particle in such an expanding field is $\mathcal{E}_{He}\sim e Z_{He} B_{ch}R_{ch}$.  

Let us assume that the electron current consists of all the electrons that initially 
were in the volume $\pi R_{ch}^3$, which is  the volume of a cavity left 
in the wake of a laser pulse, ensuring that the fields of electrons and ions are compensated 
outside the cavity and the strong fields exist only inside. 
As the electrons move through the charged ion background, they experience a plasma 
lensing effect \cite{plasma lens}, i.e., they are pinched towards the central axis 
and the radius of the electron beam is determined from the balance of the transverse 
electric field of the ion column, $E_i=2\pi e n_e R_b$, which in the reference 
frame of an e-beam is $E_i^\prime=\gamma_e E_i$, and the self field of an electron beam, 
which is equal to $E_e^\prime=2\pi e n_e^\prime R_b$. From the condition 
$E_i^\prime=E_e^\prime$ we obtain $R_b=R_{ch}/\gamma_e$. Here $\gamma_e$ 
is the Lorentz factor of the bulk of electrons accelerated forward. 
In the regime of laser pulse interaction with a NCD plasma, the 
electrons are continuously injected in the cavity behind the laser pulse, 
leading to the formation of an electron current, which is dominated by low energy electrons.

The characteristic energy of these electrons is the injection energy, 
which can be obtained from the condition on the electron velocity to be equal to the 
group velocity of an EM pulse propagating in a waveguide of radius $R_{ch}$: $
\gamma_e=\gamma_g^{ch}$ (\ref{gamma_g^ch}).        
The magnetic field, generated by these electrons, is $B_{ch}(R_b)=2\pi e n_e R_{ch}\gamma_e^2$ at $r=R_b$. Then the energy gain of an ion in the electric field at the back of the target is
\begin{equation}\label{MVA energy gain}
\mathcal{E}_{He}=m_e\,2\pi^2 Z_{He}\left(\frac{n_e}{n_{cr}}\right)\left(\frac{R_{ch}}{\lambda}\right)^4,
\end{equation}
which gives an upper estimate for the ion energy gain in the MVA regime. For example, for a 1 PW laser pulse and $n_e=n_{cr}$, the protons energy is $\mathcal{E}_{H}\approx 500$ MeV. From Eq. (\ref{MVA energy gain}), we see that the ion energy scales with the laser pulse power as $P^{2/3}$. The scaling was derived assuming the optimal match between the laser pulse and the plasma, i.e., the target length is equal to the laser pulse depletion length. We should mention here that the MVA regime is relevant for ion acceleration from targets with densities not far from  critical \cite{MVA}. For lower densities, laser pulse filamentation and Langmuir wave generation prevent efficient channel generation. At higher densities other mechanisms of laser ion acceleration come into play, and the target thickness, which is equal to the depletion length, is of the order the pulse length or smaller, which is outside the regime of MVA  applicability.


\begin{thebibliography}{99}
\bibitem{Review} E. Esarey, C. B. Schroeder, and W. P. Leemans, Rev. Mod. Phys. \textbf{81}, 1229 (2009).

\bibitem{BELLA}  W. P. Leemans, R. Duarte, E. Esarey, S. Fournier, C. G. R. Geddes, D. Lockhart, C. B. Schroeder, C. Toth, J. L. Vay, and S. Zimmermann, AIP Conf. Proc. \textbf{1299}, 3 (2010).

\bibitem{ELI} \textit{ELI-Extreme Light Infrastructure Science and Technology with Ultra-Intense Lasers WHITEBOOK}, edited by G. A. Mourou, G. Korn, W. Sandner, and J. L. Collier (THOSS Media GmbH, Berlin, 2011).

\bibitem{review} G. Mourou, T. Tajima, and S. V. Bulanov, Rev. Mod. Phys \textbf{78}, 309 (2006); H. Daido, M. Nishiuchi, and A. S. Pirozhkov, Rep. Prog. Phys. \textbf{75}, 056401 (2012); A. Macchi, M. Borghesi, and M. Passoni, Rev. Mod. Phys. \textbf{85}, 751 (2013).

\bibitem{FI} M. Roth, T. E. Cowan, M. H. Key, S. P. Hatchett, C. Brown,
W. Fountain, J. Johnson, D. M. Pennington, R. A. Snavely, S. C. Wilks, K. Yasuike,
H. Ruhl, F. Pegoraro, S. V. Bulanov, E. M. Campbell, M. D. Perry, and H. Powell,
Phys. Rev. Lett. 86, 436 (2001); V. Yu. Bychenkov, W. Rozmus, A. Maksimchuk,
D. Umstadter and C. E. Capjack, Plasma Phys. Rep. 27, 1017 (2001);
A. Macchi, A. Antonicci, S. Atzeni, D. Batani, F. Califano, F. Cornolti,
J. J. Honrubia, T. V. Lisseikina, F. Pegoraro, and M. Temporal, Nucl. Fusion 43, 362 (2003);
J. J. Honrubia, J. C. Fernandez, M. Temporal, B. M. Hegelich, and J. Meyer-ter-Vehn,
Physics of Plasmas 16, 102701 (2009);
S. Yu. Guskov, Plasma Phys. Rep. \textbf{39}, 1 (2013).

\bibitem{hadron therapy} S. V. Bulanov and V. S. Khoroshkov, Plasma. Phys. Rep. \textbf{28}, 453
(2002).

\bibitem{radiography} M. Borghesi, J. Fuchs, S. V. Bulanov, A. J. Mackinnon,
P. K. Patel, and M. Roth, Fusion Science and Technology \textbf{49}, 412
(2006).

\bibitem{injection} K. Krushelnick, E. L. Clark, R. Allott, F. N. Beg,
C. N. Danson, A. Machacek, V. Malka, Z. Najmudin, D. Neely, P. A.
Norreys, M. R. Salvati, M. I. K. Santala, M. Tatarakis, I. Watts, M.
Zepf, A. E. Dangor, Plasma Science, IEEE Transactions on 28, 1184 -
1189 (2000).

\bibitem{nuclear} M. Nishiuchi, H. Sakaki, K. Nishio, R. Orlandi, H. Sako, 
T. A. Pikuz, A. Ya. Faenov, T. Zh. Esirkepov, A. S. Pirozhkov, K. Matsukawa, 
A. Sagisaka, K. Ogura, M. Kanasaki1, H. Kiriyama, Y. Fukuda, H. Koura, M. Kando, 
T. Yamauchi, Y. Watanabe, S. V. Bulanov, K. Kondo, K. Imai, and S. Nagamiya, arXiv:1402.5729;
N. V. Zamfir, Eur. Phys. J. Special Topics 223, 1221 (2014).

\bibitem{MeV protons} T. Zh. Esirkepov, 
%\textit{et al.}, 
Y. Sentoku, K. Mima, K. Nishihara, F. Califano, F. Pegoraro, N. M. Naumova, S. V. Bulanov, Y. Ueshima, T. V. Liseikina, V. A. Vshivkov, and Y. Kato,
JETP Lett. \textbf{70}, 82 (1999); 
A. M. Pukhov, Phys. Rev. Lett. \textbf{86}, 3562 (2001); 
Y. Sentoku, 
%\textit{et al.}, 
V.Y. Bychenkov, K. Flippo, A. Maksimchuk, K. Mima, G. Mourou, Z.M. Sheng, and D. Umstadter,
Appl. Phys. B \textbf{74}, 207 (2002); 
A. J. Mackinnon, Y. Sentoku, P. K. Patel, D. W. Price, S. Hatchett, M. H. Key, C. Andersen, R. Snavely, and R. R. Freeman, Phys. Rev. Lett. \textbf{88}, 215006 (2002);
S. V. Bulanov, 
%\textit{et al.}, 
T. Zh. Esirkepov, F. Califano, Y. Kato, T. V. Liseikina, K. Mima, N. M. Naumova, K. Nishihara, F. Pegoraro, H. Ruhl, Y. Sentoku, and Y. Ueshima,
JETP Lett. \textbf{71}, 407 (2000); 
Y. Sentoku, 
%\textit{et al.}, 
T. V. Liseikina, T. Zh. Esirkepov, F. Califano, N. M. Naumova, Y. Ueshima, V. A. Vshivkov, Y. Kato, K. Mima, K. Nishihara, F. Pegoraro, and S. V. Bulanov,
Phys. Rev. E \textbf{62}, 7271 (2000); 
H. Ruhl, S. V. Bulanov, T. E. Cowan, T. V. Liseikina, P. Nickles, F. Pegoraro, M. Roth, W. Sandner, Plasma Phys. Rep.
\textbf{27}, 411 (2001).

\bibitem{TeV protons} L. Yu, H. Xu, W. M. Wang, Z. M. Sheng, B. F. Shen, W. Yu, and J. Zhang, New J. Phys. \textbf{12}, 045021 (2010); X. Zhang, B. Shen, L. Ji, F. Wang, M. Wen, W. Wang, J. Xu, and Y. Yu, Phys. Plasmas \textbf{17}, 123102 (2010).

\bibitem{slow wave_PRL} S. S. Bulanov, 
%{\it et al.},
E. Esarey, C. B. Schroeder, S. V. Bulanov, T. Zh. Esirkepov, M. Kando, F. Pegoraro, and W. P. Leemans,
Phys. Rev. Lett. \textbf{114}, 105003 (2015).

\bibitem{TNSA} S. C. Wilks, A. B. Langdon, T. E. Cowan, M. Roth, M. Singh, S. Hatchett, M. H. Key, D. Pennington, A. MacKinnon, and R. A. Snavely, Phys. Plasmas \textbf{8}, 542 (2001).

\bibitem{CE} I. Last, I. Scheck, and J. Jortner, J. Chem. Phys. \textbf{107}, 6685 (1997); 
S. V. Bulanov, T. Zh. Esirkepov, V. S. Khoroshkov, A. V. Kuznetsov and F. Pegoraro, Phys. Lett. A 
\textbf{299}, 240 (2002); 
V. F. Kovalev and V. Yu. Bychenkov, Phys. Rev. Lett. \textbf{90}, 185004 (2003); 
E. Fourkal, I. Velchev, and C.-M. Ma, Phys. Rev. E \textbf{71}, 036421 (2005); 
I. Last and J. Jortner, Proc. Natl. Acad. Sci. U.S.A. \textbf{102}, 1291 (2005);
M. Murakami and M. M. Basko, Phys. Plasmas \textbf{13}, 012105 (2006). 

\bibitem{RPA} T. Esirkepov, M. Borghesi, S. V. Bulanov, G. Mourou, and T. Tajima, Phys. Rev. Lett. \textbf{92}, 175003 (2004).

\bibitem{MVA_first} A. V. Kuznetsov, T. Zh. Esirkepov, F. F. Kamenets, and S. V. Bulanov, Fiz. Plazmy 27, 225 (2001) [Plasma Phys. Rep. 27, 211 (2001)]; Y. Sentoku, T. V. Liseikina, T. Zh. Esirkepov, F. Califano, N. M. Naumova, Y. Ueshima, V. A. Vshivkov, Y. Kato, K. Mima, K. Nishihara, F. Pegoraro, and S. V. Bulanov, Phys. Rev. E 62, 7271 - 7281 (2000); K. Matsukado, T. Esirkepov, K. Kinoshita, H. Daido, T. Utsumi, Z. Li, A. Fukumi, Y. Hayashi, S. Orimo, M. Nishiuchi, S. V. Bulanov, T. Tajima, A. Noda, Y. Iwashita, T. Shirai, T. Takeuchi, S. Nakamura, A. Yamazaki, M. Ikegami, T. Mihara, A. Morita, M. Uesaka, K. Yoshii, T. Watanabe, T. Hosokai, A. Zhidkov, A. Ogata, Y. Wada, and T. Kubota, Phys. Rev. Lett. 91, 215001 (2003).

\bibitem{MVA} S. S. Bulanov, V. Yu. Bychenkov, V. Chvykov, G. Kalinchenko, D. W. Litzenberg, T. Matsuoka, A. G. R. Thomas, L. Willingale, V. Yanovsky, K. Krushelnick, and A. Maksimchuck, Phys. Plasmas \textbf{17}, 1 (2010).

\bibitem{MVA_new} M. H. Helle, D. F. Gordon, D. Kaganovich, Y.-H. Chen, and A. Ting, Proc. SPIE  9514, 951409 (2015). 

\bibitem{SWA} D. Haberberger, S. Tochitsky, F. Fiuza, \textit{et al.}, Nature Phys. \textbf{8}, 95 (2012); F. Fiuza,
A. Stockem, E. Boella, \textit{et al.}, Phys. Rev. Lett. \textbf{109}, 215001 (2012); A. Macchi, A. S. Nindrayong, F. Pegoraro, Phys. Rev. E \textbf{85}, 046402 (2012).

\bibitem{TNSA_exp} S. A. Gaillard, T. Kluge, K. A. Flippo, M. Bussmann, B. Gall, T. Lockard, M. Geissel, D. T. Offermann, M. Schollmeier, Y. Sentoku, and T. E. Cowan, Phys. Plasmas \textbf{18}, 056710 (2011). 

\bibitem{RM review}S. V. Bulanov, T. Zh. Esirkepov, M. Kando, A. S. Pirozhkov, and N. N. Rosanov, Physics - Uspekhi \textbf{56}, 429 (2013).

\bibitem{hole-boring RPA} S. Wilks, W. Kruer, M. Tabak, and A. B. Langdon, Phys. Rev. Lett. \textbf{69}, 1383 (1992); 
N. M. Naumova, T. Schlegel, V. T. Tikhonchuk, C. Labaune, I. V. Sokolov, and G. Mourou, Phys. Rev. Lett. \textbf{102}, 025002 (2009). 

\bibitem{Einstein} A. Einstein, Ann. Phys. (Leipzig) {\bf 17} (1905) 891.

\bibitem{COMREN}S. V. Bulanov, T. Zh. Esirkepov, F. Pegoraro, M. Borghesi, Comptes Rendus Physique {\bf 10}, 216 (2009).

\bibitem{RPA_recent} O. Klimo, J. Psikal, J. Limpouch, and V. T. Tikhonchuk, Phys. Rev. ST Accel. Beams {\bf 11}, 031301 (2008); A. P. L. Robinson, M. Zepf, S. Kar, R. G. Evans, and C. Bellei, New J. Phys. {\bf 10}, 013021 (2008); B. Qiao, M. Zepf, M. Borghesi, and M. Geissler, Phys. Rev. Lett. {\bf 102}, 145002 (2009); X. Q. Yan, C. Lin,Z. M. Sheng, Z. Y. Guo, B. C. Liu, Y. R. Lu, J. X. Fang, and J. E. Chen, Phys. Rev. Lett. {\bf 103}, 135001 (2009); J.-L. Liu, M. Chen, J. Zheng, Z.-M. Sheng, C.-S. Liu, Phys. Plasmas {\bf 20}, 063107 (2013).

\bibitem{RPA exp} S. Kar, M. Borghesi, S. V. Bulanov, M. H. Key, T. V. Liseykina, A. Macchi, A. J. Mackinnon, P. K. Patel, L. Romagnani, A. Schiavi, O. Willi, Phys. Rev. Lett,  \textbf{100}, 225004 (2008); K. U. Akli, S. B. Hansen, A. J. Kemp, R. R. Freeman, F. N. Beg, D. C. Clark, S. D. Chen, D. Hey, S. P. Hatchett, K. Highbarger, E. Giraldez, J. S. Green, G. Gregori, K. L. Lancaster, T. Ma, A. J. MacKinnon, P. Norreys, N. Patel, J. Pasley, C. Shearer, R. B. Stephens, C. Stoeckl, M. Storm, W. Theobald, L. D. Van Woerkom, R. Weber, M. H. Key, Phys. Rev. Lett. \textbf{100}, 165002 (2008); A. Henig, S. Steinke, M. Schnurer, T. Sokollik, R. Horlein, D. Kiefer, D. Jung, J. Schreiber, B. M. Hegelich, X. Q. Yan, J. Meyer-ter-Vehn, T. Tajima, P. V. Nickles, W. Sandner, and D. Habs, Phys. Rev. Lett. \textbf{103}, 245003 (2009); C. A. J. Palmer, N. P. Dover, I. Pogorelsky, M. Babzien, G. I. Dudnikova, M. Ispiriyan, M. N. Polyanskiy, J. Schreiber, P. Shkolnikov, V. Yakimenko, and Z. Najmudin, Phys. Rev. Lett. \textbf{106}, 014801 (2011); S. Steinke, P. Hilz, M. Schnurer, G. Priebe, J. Branzel, F. Abicht, D. Kiefer, C. Kreuzer, T. Ostermayr, J. Schreiber, A. A. Andreev, T. P. Yu, A. Pukhov, and W. Sandner, Phys. Rev. ST Accel. Beams \textbf{16}, 011303 (2013).

\bibitem{Willingale2006} L. Willingale, S. P. D. Mangles, P. M. Nilson, R. J. Clarke, A. E. Dangor, M. C. Kaluza, S. Karsch, K. L. Lancaster, W. B. Mori, Z. Najmudin, J. Schreiber, A. G. R. Thomas, M. S. Wei, and K. Krushelnick, Phys. Rev. Lett. \textbf{96}, 245002 (2006).

\bibitem{Willingale2011} L. Willingale, P. M. Nilson, A. G. R. Thomas, S. S. Bulanov, Maksimchuk, W. Nazarov, T. C. Sangster, C. Stoeck, and K. Krushelnick, Phys. Plasmas \textbf{18} 056706 (2011).

\bibitem{Fukuda} Y. Fukuda, A. Ya. Faenov, M. Tampo, T. A. Pikuz, T. Nakamura, M. Kando, Y. Hayashi, A. Yogo, H. Sakaki, T. Kameshima, A. S. Pirozhkov, K. Ogura, M. Mori, T. Zh. Esirkepov, J. Koga, A. S. Boldarev, V. A. Gasilov, A. I. Magunov, T. Yamauchi, R. Kodama, P. R. Bolton, Y. Kato, T. Tajima, H. Daido, and S. V. Bulanov, Phys. Rev. Lett. \textbf{103}, 165002 (2009).

\bibitem{slow wave} S. V. Bulanov, T. Zh. Esirkepov, M. Kando, F. Pegoraro, S. S. Bulanov,  C. G. R. Geddes, C. B. Schroeder, E. Esarey, W. P. Leemans, Phys. Plasmas, \textbf{19}, 103105 (2012); A. A. Sahai and T. C. Katsouleas, arXiv:1411.2401.

\bibitem{BOA} L. Yin, B. J. Albright, B. M. Hegelich, K. J. Bowers, K. A. Flippo, T. J. T. Kwan, and J. C. Fernandez, Phys. Plasmas \textbf{14}, 056706 (2007).

\bibitem{DCE} S. S. Bulanov, A. Brantov, V. Yu. Bychenkov, V. Chvykov, G. Kalinchenko, T. Matsuoka, P. Rousseau, S. Reed, V. Yanovsky, D. W. Litzenberg, and A. Maksimchuk, Med. Phys. \textbf{35}, 1770 (2008); 
S. S. Bulanov, A. Brantov, V. Yu. Bychenkov, V. Chvykov, G. Kalinchenko, T. Matsuoka, P. Rousseau, V. Yanovsky, D. W. Litzenberg, K. Krushelnick, and A. Maksimchuk, Phys. Rev. E \textbf{78}, 026412 (2008).

\bibitem{transparency} S. Palaniyappan, B. M. Hegelich, H.-C. Wu, D. Jung, D. C. Gautier, L. Yin, B. J. Albright, R. P. Johnson, T. Shimada, S. Letzring, D. T. Offermann, J. Ren, C. Huang, R. Horlein, B. Dromey, J. C. Fernandez, and R. C. Shah, Nature Physics \textbf{8}, 763 (2012);
B. M. Hegelich, 
%{\it et al.},
I. Pomerantz, L. Yin, H. C. Wu, D. Jung, B. J. Albright, D. C. Gautier, S. Letzring, S. Palaniyappan, R. Shah, K. Allinger, R. Horlein, J. Schreiber, D. Habs, J. Blakeney, G. Dyer, L. Fuller, E. Gaul, E. Mccary, A. R. Meadows, C. Wang, T. Ditmire, and J. C. Fernandez, New J. Phys. \textbf{15}, 085015 (2013);
D. Jung,
%{\it et al.}, 
B. J. Albright, L. Yin, D. C. Gautier, R. Shah, S. Palaniyappan, S. Letzring, B. Dromey, H.-C. Wu, T. Shimada, R. P. Johnson, M. Roth, J. C. Fernandez, D. Habs, and B. M. Hegelich, New J. Phys. \textbf{15}, 123035 (2013);


\bibitem{optimized RPA} S. S. Bulanov, C. B. Schroeder, E. Esarey, and W. P. Leemans, Phys. Plasmas \textbf{19}, 093112 (2012).

\bibitem{Macchi} A. Sgattoni, P. Londrillo, A. Macchi, and M. Passoni, Phys. Rev. E \textbf{85}, 036405 (2012).

\bibitem{Zepf} H. Y. Wang, X. Q. Yan, and M. Zepf, Phys. Rev. ST Accel. Beams \textbf{18}, 021302 (2015). 

\bibitem{Dollar} F. Dollar, C. Zulick, A. G. R. Thomas, V. Chvykov, J. Davis, G. Kalinchenko, T. Matsuoka, C. McGuffey, G. M. Petrov, L. Willingale, V. Yanovsky, A. Maksimchuk, and K. Krushelnick, Phys. Rev. Lett. \textbf{108}, 175005 (2012). 

\bibitem{unlimited} S. V. Bulanov, E. Yu. Echkina, T. Zh. Esirkepov, I. N. Inovenkov, M. Kando, F. Pegoraro, and G. Korn, Phys. Rev. Lett. \textbf{104}, 135003 (2010); Phys. Plasmas \textbf{17}, 063102 (2010).

\bibitem{Vshivkov}  V. A. Vshivkov, N. M. Naumova, F. Pegoraro, and S. V. Bulanov, Phys. Plasmas \textbf{5}, 2727 (1998).

\bibitem{optimal} T. Esirkepov, M. Yamagiwa, and T. Tajima, Phys. Rev. Lett. 96, 105001 (2006).

\bibitem{REMP} T. Zh. Esirkepov, Comput. Phys. Comm. \textbf{135}, 144 (2001).

\bibitem{group velocity} E. Esarey, P. Sprangle, M. Pilloff, J. Krall, J. Opt. Soc. Amer. B \textbf{12}, 1695 (1995).

\bibitem{plasma lens} P. Chen,  
%{\it et al.},
J. J. Su, T. Katsouleas, S. Wilks, and J. M. Dawson, Plasma Science, 
IEEE Transactions on \textbf{PS-15}, 218 (1987).

\bibitem{Koga_photon_photon} J. K. Koga, S. V. Bulanov, T. Zh. Esirkepov, A. S. Pirozhkov, and M. Kando, Phys. Rev. A \textbf{86}, 053823 (2012).

\bibitem{ShFr} S. V. Bulanov, I. N. Inovenkov, V. I. Kirsanov, N. M. Naumova, and A. S. Sakharov, 
%"Ultra fast depletion of ultra-short and relativistically strong laser pulses in an underdense plasma", 
Physics of Fluids B {\bf 4}, 1935 (1992);
S. V. Bulanov, I. N. Inovenkov, V. I. Kirsanov, N. M. Naumova, A. S. Sakharov, and H. A. Shakh, 
%"Stationary shock-front of a relativistically strong electromagnetic radiation in an underdense plasma", 
Physica Scripta {\bf 47}, 209 (1993); 
C. D. Decker, W. B. Mori, K.C. Tzeng, and T. Katsouleas,
%"The evolution of ultraintense, shortpulse lasers in underdense plasmas",
Phys. Plasmas {\bf 3}, 2047 (1996).


\bibitem{helium} S. S. Bulanov,
%{\it et al.},
E. Esarey, C. B. Schroeder, W. P. Leemans, S. V. Bulanov, D. Margarone, G. Korn, and T. Haberer, Phys. Rev. ST Accel. Beams \textbf{18}, 061302 (2015).

\end{thebibliography}
\end{document}